\documentclass{jfm}
\usepackage{graphicx}
\usepackage{epstopdf, epsfig}
\usepackage{mathrsfs}
\usepackage{color}
\usepackage[left]{lineno}

\nolinenumbers

\shorttitle{Mean Flow and Turbulence in Unsteady Canopy Layers}
\shortauthor{W. Li and M. G. Giometto}

\title{Mean Flow and Turbulence in Unsteady Canopy Layers}

\author{Weiyi Li\aff{1}
 \and Marco G. Giometto\aff{1}
 \corresp{\email{mg3929@columbia.edu}}}

\affiliation{\aff{1}Department of Civil Engineering and Engineering Mechanics, Columbia University, New York, NY 10027}

\begin{document}

\maketitle

\begin{abstract}

Non-stationarity is the rule in the atmospheric boundary layer (ABL). Under such conditions, the flow may experience departures from equilibrium with the underlying surface stress, misalignment of shear stresses and strain rates, and three-dimensionality in turbulence statistics.
Existing ABL flow theories are primarily established for statistically stationary flow conditions and cannot predict such behaviors.
Motivated by this knowledge gap, this study analyzes the impact of time-varying pressure gradients on mean flow and turbulence over urban-like surfaces.
A series of large-eddy simulations of pulsatile flow over cuboid arrays is performed, programmatically varying the oscillation amplitude $\alpha$ and forcing frequency $\omega$.
The analysis focuses on both longtime-averaged and phase-dependent flow dynamics.
Inspection of longtime-averaged velocity profiles reveals that the aerodynamic roughness length $z_0$ increases with $\alpha$ and $\omega$, whereas the displacement height $d$ appears to be insensitive to these parameters.
In terms of phase-averaged flow statistics, it is found that $\alpha$ primarily controls the oscillation amplitude of the streamwise velocity and Reynolds stresses, but has a negligible impact on their wall-normal structure.
On the other hand, $\omega$ determines the size of the region affected by the unsteady forcing, which identifies the so-called Stokes layer thickness $\delta_s$.
Within the Stokes layer, phase-averaged resolved Reynolds stress profiles feature substantial variations during the pulsatile cycle, and the turbulence is out of equilibrium with the mean flow.
Two phenomenological models have been proposed that capture the influence of flow unsteadiness on $z_0$ and $\delta_s$, respectively.

\end{abstract}

\section{Introduction}\label{sec:intro}
Advancing our conceptual understanding and ability to predictively model exchange processes between urban areas and the atmosphere is of critical importance to a wide range of applications, including urban air quality control \citep{britter2003flow,barlow2004scalar,pascheke2008wind}, urban microclimate studies \citep{roth2012urban,li2014quality,ramamurthy2017high}, and weather and climate forecasting \citep{holtslag2013stable}, to name but a few.
It hence comes as no surprise that substantial efforts have been devoted towards this goal over the past decades, via, e.g., numerical simulations \citep{bou2004large,xie2008large,cheng2015adjustment,giometto2016spatial,sadique2017aerodynamic,auvinen2017numerical,zhu2017parametric,li2019contrasts}, wind tunnel experiments \citep{raupach1980wind,bohm2013turbulence,marucci2020stable}, and observational studies \citep{rotach1993turbulence,kastner2004mean,rotach2005bubble, Christen2009}.
These studies have explored the functional dependence of flow statistics on urban canopy geometry \citep{lettau1969note,raupach1992drag,macdonald1998improved,coceal2004canopy,yang2016exponential,li2022bridging}, characterized the topology of coherent structures \citep{kanda2004large,coceal2007structure,Christen2007,li2011coherent,inagaki2012large}, and derived scaling laws for scalar transfer between the urban canopy and the atmosphere \citep{pascheke2008wind,cheng2016large,li2019contrasts}, amongst others.
Most of the previous works have focused on atmospheric boundary layer (ABL) flow under (quasi-)stationary conditions.
However, stationarity is of rare occurrence in the ABL \citep{mahrt2020non}, and theories based on equilibrium turbulence are therefore often unable to grasp the full range of physics characterizing ABL flow environments.

Major drivers of non-stationarity in the ABL include time-varying horizontal pressure gradients, associated with non-turbulent motions ranging from submeso to synoptic scales, and time-dependent thermal forcings, induced by the diurnal cycle or by cloud-induced time variations of the incoming solar radiation \citep{mahrt2020non}.
These conditions often result in departures from equilibrium turbulence, with important implications on time- and area-averaged exchange processes between the land surface and the atmosphere.
The first kind of non-stationarity was examined in \cite{mahrt2007influence,mahrt2008influence,mahrt2013non}, which showed that time-variations of the driving pressure gradient can enhance momentum transport under strong stable atmospheric stratifications.
The second kind of non-stationarity was instead analyzed in \cite{hicks2018relevance} making use of data from different field campaigns, and showed that the surface heat flux can change so rapidly during the morning and late afternoon transition that the relations for equilibrium turbulence no longer hold. 

Numerical studies have also been recently conducted to study how exchange processes between the land surface and the atmosphere are modulated by non-stationarity in the ABL.
In their study, Edwards \cite{edwards2006simulation} conducted a comparison of a prevailing single-column model based on equilibrium turbulence theories with the observations of an evening transition ABL, as well as results from large-eddy simulations (LES). Their findings emphasized the inadequacy of equilibrium turbulence theories in capturing the complex behavior of ABL flows during rapid changes in thermal surface forcing. 
This breakdown of equilibrium turbulence was particularly notable during the evening transition period, which is known for its rapid changes in thermal forcing.
\cite{momen2017mean} investigated the response of the Ekman boundary layer to oscillating pressure gradients, and found that quasi-equilibrium turbulence is maintained only when the oscillation period is much larger than the characteristic time scale of the turbulence.
The majority of the efforts have focused on atmospheric boundary layer flow over modeled roughness, where the flow dynamics in the roughness sublayer---that layer of the atmosphere that extends from the ground surface up to about 2-5 times the mean height of roughness elements \citep{fernando2010fluid}---are bypassed, and surface drag is usually evaluated via an equilibrium wall-layer model (see, e.g., \cite{momen2017mean}), irrespective of the equilibrium-theory limitations outlined above.
It hence remains unclear how unsteadiness impacts flow statistics and the structure of atmospheric turbulence in the roughness sublayer.
Roughness sublayer flow directly controls exchanges of mass, energy, and momentum between the land surface and the atmosphere, and understanding the dependence of flow statistics and structural changes in the turbulence topology on flow unsteadiness is hence important in order to advance our ability to understand and predictively model these flow processes.

This study contributes to addressing this knowledge gap by focusing on non-stationarity roughness sublayer flow induced by time-varying pressure gradients.
Unsteady pressure gradients in the real-world ABL can be characterized by periodic and aperiodic variations in both magnitude and direction.
In this study, we limit our attention to a pulsatile streamwise pressure-gradient forcing, consisting of a constant mean and a sinusoidal oscillating component. This approach has two major merits.
First, the temporal evolution of flow dynamics and associated statistics, as well as structural changes in turbulence, can be easily characterized thanks to the time-periodic nature of the flow unsteadiness.
Second, the time scale of the pulsatile forcing is well defined, and can hence be varied programmatically to encompass a range of representative flow regimes.

Pulsatile turbulent flows over aerodynamically smooth surfaces have been the subject of active research in the mechanical engineering community because of their relevance across a range of applications; these include industrial (e.g., a rotating or poppet valve) and biological (blood in arteries) flows.
The corresponding laminar solution is an extension of Stokes's second problem \citep{stokes1901effect}, where the modulation of the flow field by unsteady pressure gradient is confined to a layer of finite thickness known as the ``Stokes layer''.
The thickness of the Stokes layer $\delta_s$ is a function of the pulsatile forcing frequency $\omega$, i.e., $\delta_s=2l_s$, where $l_s=\sqrt{2\nu/\omega}$ is the so-called Stokes length scale and $\nu$ is the kinematic viscosity of the fluid.
In the turbulent flow regime, it has been found that the characteristics of the pulsatile flow are not only dependent on the forcing frequency, but also on the amplitude of the oscillation.
Substantial efforts have been devoted to investigating this problem, from both an experimental \citep{ramaprian1980experimental,tu1983fully,ramaprian1983fully,mao1986studies,brereton1990response,tardu1993wall,Tardu2005} and a computational perspective \citep{scotti2001numerical,manna2012pulsating,manna2015pulsating,weng2016numerical}.
\cite{scotti2001numerical} drew an analogy to the Stokes length and proposed a turbulent Stokes length scale, which can be expressed in inner units as
\begin{equation}
    l_t^+=\frac{\overline{u}_\tau}{\nu}\left( \frac{2(\nu+\nu_t)}{\omega}\right)^{1/2}\ ,
    \label{eq:lt_pio_init}
\end{equation}
where $\overline{u}_\tau$ is the friction velocity based on the surface friction averaged over pulsatile cycles and $\nu_t$ is the so-called eddy viscosity.
Parameterizing the eddy viscosity in terms of the Stokes turbulent length scale, i.e., $\nu_t = \kappa \overline{u}_\tau l_t$, where $\kappa$ is the von K\'{a}rm\'{a}n constant, and substituting into (\ref{eq:lt_pio_init}) one obtains
\begin{equation}
    l_t^+ = l_s^+\left(\frac{\kappa l_s^+}{2}+\left(1+\left( \frac{\kappa l_s^+}{2}\right)^{\frac{1}{2}}\right) \right)\ .    \label{eq:lt_pio}
\end{equation}

When $l_t^+$ is large (e.g., when $\omega \rightarrow 0$), the flow is in a quasi-steady state.
Under such a condition, the flow at each pulsatile phase resembles a statistically stationary boundary layer flow, provided that the instantaneous friction velocity is used to normalize statistics.
As $\omega$ increases and $l_t^+$ becomes of the order of the open channel height $L_3^+$, the entire flow is affected by the pulsation, i.e., time lags occur between flow statistics at different elevations, and turbulence undergoes substantial structural changes from its equilibrium configuration.
When $l_t^+<L_3^+/2$, the flow modulation induced by the pulsation is confined within the Stokes layer $\delta_s^+=2l_t^+$.
Above the Stokes layer, one can observe a plug-flow region, with the turbulence being frozen to its equilibrium configuration and simply advected by the mean flow pulsation.
A few years later, \cite{bhaganagar2008direct} conducted a series of direct numerical simulations (DNS) of low-Reynolds-number pulsatile flow over transitionally rough surfaces.
She found that flow responses to pulsatile forcing are generally similar to those in smooth-wall cases, when the roughness size is of the same order of magnitude as the viscous sublayer thickness. 
The only exception is that, as the pulsation frequency approaches the frequency of vortex shedding from the roughness elements, the longtime averaged velocity profile deviates significantly from that of the steady flow case due to the resonance between the pulsation and the vortex shedding.
In the context of a similar flow system, \cite{patil2022drag} also reported a comparable observation in their DNS study.

In addition to the work of \cite{bhaganagar2008direct,patil2022drag}, pulsatile flow over small-scale roughness, e.g., sand grain roughness, has also been studied extensively in the oceanic context, i.e., combined current-wave boundary layers, which play a crucial role in controlling sediment transport and associated erosion in coastal environments   \citep{grant1979combined,kemp1982interaction,myrhaug1989combined,sleath1991velocities,soulsby1993wave,mathisen1996waves,fredsoe1999wave,yang2006velocity,yuan2015experimental}.
The thickness of the wave boundary layer, which is the equivalent of the Stokes layer in the engineering community, is defined as
\begin{equation}
    \delta_w=\frac{2\kappa u_{\tau,max}}{\omega}\ ,
    \label{eq:lw}
\end{equation}
where $u_{\tau,max}=\sqrt{\tau_{max}}$, and $\tau_{max}$ denotes the maximum of kinematic shear stress at the surface during the pulsatile cycle.
Within the wave boundary layer, mean flow and turbulence are controlled by the nonlinear interaction between currents and waves.
Above this region, the modulation of turbulence by waves vanishes.
The wall-normal distribution of the averaged velocity over the pulsatile cycle deviates from the classic logarithmic profile, and is characterized by a ``two-log" profile, i.e., the velocity exhibits a logarithmic profile with the actual roughness length within the wave boundary layer and a different one characterized by a larger roughness length further aloft \citep{grant1986continental,fredsoe1999wave,yang2006velocity,yuan2015experimental}.
Such behavior was first predicted by a two-layer time-invariant eddy viscosity model by \cite{grant1979combined}, followed by many variants and improvements \citep{myrhaug1989combined,sleath1991velocities,yuan2015experimental}.

On the contrary, pulsatile flow at high Reynolds numbers over large roughness elements, such as buildings, has received far less attention.
\cite{yu2022boundary} conducted a series of LES of combined wave-current flows over arrays of hemispheres, which can be seen as a surrogate of reefs near the coastal ocean. They focused only on low Keulegan-Carpenter numbers $KC\sim \mathcal{O}(1-10)$, where $KC$ is defined as the ratio between the wave excursion $U_w T$ and the diameter of the hemispheres, and $U_w$ and $T$ are the wave orbital velocity and the wave period, respectively \citep{keulegan1958forces}.
However, conclusions from \cite{yu2022boundary} cannot be readily applied to pulsatile flow over urban-like roughness (i.e., large obstacles with sharp edges), mainly because different surface morphologies yield distinct air-canopy interaction regimes under pulsatile forcings \citep{carr2017surface}. 

Motivated by this knowledge gap, this study proposes a detailed analysis on the dynamics of the mean flow and turbulence in high-Reynolds-number pulsatile flow over idealized urban canopies.
The analysis is carried out based on a series of LES of pulsatile flow past an array of surface-mounted cuboids, where the frequency and amplitude of the pressure gradient are programmatically varied.
The LES technique has been shown capable of capturing the major flow features of pulsatile flow over various surface conditions \citep{scotti2001numerical,chang2004modeling}.
The objective of this study is to answer fundamental questions pertaining to the impacts of the considered flow unsteadiness on the mean flow and turbulence in the urban boundary layer:
\begin{enumerate}
  \item Does the presence of flow unsteadiness alter the mean flow profile in a longtime-averaged sense? If so, how do such modifications reflect in the aerodynamic surface parameters?
  \item To what extent does the unsteady pressure gradient impact the overall momentum transport and turbulence generation in a longtime-averaged sense within and above the canopy?
  \item How do the phase-averaged mean flow and turbulence behave in response to the periodically varying pressure gradient? How are such phase-dependent behaviors controlled by the oscillation amplitude and the forcing frequency?
\end{enumerate}

This paper is organized as follows.
Section~\ref{sec:meth} introduces the numerical algorithm and the setup of simulations, along with the flow decomposition and averaging procedure.
Results are presented and discussed in \S\ref{sec:results}.
Concluding remarks are given in \S\ref{sec:conclusion}.

\section{Methodology}\label{sec:meth}

\subsection{Numerical procedure}\label{subsec:num}

A suite of LES is performed using an extensively validated in-house code \citep{albertson1999natural,albertson1999surface,bou2005scale,chamecki2009large,anderson2015numerical,fang2015large,li2016quality,giometto2016spatial}.
The code solves the filtered continuity and momentum transport equations in a Cartesian reference system, which read
\begin{equation}
    \frac{\partial u_i}{\partial x_i}=0\ ,
    \label{eq:continuity}
\end{equation}
\begin{equation}
    \frac{\partial u_i}{\partial t} +  u_j (\frac{\partial u_i}{\partial x_j}-\frac{\partial u_j}{\partial x_i})  = - \frac{1}{\rho} \frac{\partial p^*}{ \partial  x_i} - \frac{\partial \tau_{ij}}{\partial x_j} - \frac{1}{\rho }\frac{\partial p_\infty}{ \partial x_1} \delta_{i1}+F_i\ ,
    \label{eq:momentum}
\end{equation}
where $u_1$, $u_2$, and $u_3$ are the filtered velocities along the streamwise $(x_1)$, lateral $(x_2)$, and wall-normal $(x_3)$ direction, respectively.
The advection term is written in the rotational form to ensure kinetic energy conservation in the discrete sense \citep{orszag1975numerical}.
$\rho$ represents the constant fluid density, $\tau_{ij}$ is the deviatoric component of the subgrid-scale (SGS) stress tensor, which is evaluated via the Lagrangian scale-dependent dynamic (LASD) Smagorinsky model \citep{bou2005scale}.
The LASD model has been extensively validated in wall-modeled simulations of unsteady atmospheric boundary layer flow \citep{momen2017mean,salesky2017nature} and in the simulation of flow over surface-resolved urban-like canopies \citep{anderson2015numerical,li2016quality,giometto2016spatial,yang2016mean}.
Note that viscous stresses are neglected in the current study; this assumption is valid as the typical Reynolds number of the ABL flows is $Re\sim\mathcal{O}(10^9)$, and the flow is in the fully rough regime.
$p^*=p+\frac{1}{3}\rho \tau_{ii}+\frac{1}{2} \rho u_i u_i$ is the modified pressure, which accounts for the trace of SGS stress and resolved turbulent kinetic energy.
The flow is driven by a spatially uniform but temporally periodic pressure gradient, i.e.,
\begin{equation}
  - {\partial p_\infty}/{\partial x_1} = \rho f_m\left[1+\alpha_p \sin(\omega t)\right] \ ,
\end{equation}
where $f_m$ denotes the mean pressure gradient. $\alpha_p$ is a constant controlling the amplitude of the forcing, and $\omega$ represents the forcing frequency.
$\delta_{ij}$ is the Kronecker delta tensor.

Periodic boundary conditions apply in the wall-parallel directions, and free-slip boundary conditions are employed at the upper boundary.
The lower surface is representative of an array of uniformly distributed cuboids, which serves as a surrogate of urban landscapes.
This approach has been commonly adopted in studies of ABL, as this type of surface morphology is characterized by a limited number of length scales, making it amenable to analytical treatment \citep{reynolds2008measurements,cheng2015adjustment,tomas2016stable,basley2019structure,omidvar2020plume}. 
Such an approach is justified on the basis that one should first study a problem in its simplest setup before introducing additional complexities. 
Nonetheless, it is crucial to acknowledge that the introduction of randomness in roughness alters the flow characteristics and the generation of turbulence in rough-wall boundary layer flows.
\cite{xie2008large} studied flows over random urban-like obstacles and found that turbulence features in the roughness sublayer are controlled by the randomness in the roughness. 
\cite{giometto2016spatial} conducted an LES study and highlighted that roughness randomness enhances the dispersive stress in the roughness sublayer.
\cite{chau2012understanding}  carried out a series of DNS of flow over transitionally rough surfaces, demonstrating that different levels of roughness randomness lead to distinct turbulence structures in the near-wall region and subsequently affect turbulence intensities. 

Spatial derivatives in the wall-parallel directions are computed via a pseudo-spectral collocation method based on truncated Fourier expansions \citep{orszag1970analytical}, whereas a second-order staggered finite difference scheme is employed in the wall-normal direction.
A second-order Adams-Bashforth scheme is adopted for time integration.
Nonlinear advection terms are de-aliased via the $3/2$ rule \citep{canuto2007spectral,margairaz2018comparison}.
Roughness elements are explicitly resolved via a discrete-forcing immersed boundary method (IBM), which is also referred to as the direct forcing IBM in the engineering fluid mechanics community \citep{yusof1996interaction,mittal2005immersed,fang2011towards}.
The IBM was originally developed in \cite{yusof1996interaction} and first introduced to ABL studies by \cite{chester2007modeling}.
Since then, the IBM has been extensively validated in subsequent studies \citep[e.g.][]{graham2012modeling,cheng2015adjustment,giometto2016spatial,anderson2016amplitude,yang2018numerical,li2019contrasts}.
Specifically, an artificial force $F_i$ drives the velocity to zero within the cuboids, and an inviscid equilibrium logarithmic wall-layer model \citep{moeng1984large,giometto2016spatial} is applied over a narrow band centered at the fluid-solid interface to evaluate the wall stresses.
As shown in Appendix~\ref{sec:wall_model_justify}, for flow over cuboids in the fully rough regime, the use of an equilibrium wall model does not impact the flow field significantly. 
\cite{xie2006and} reached a similar conclusion for a comparable flow system.
The incompressibility condition is then enforced via a pressure-projection approach \cite{kim1985application}.

\begin{figure}
  \centerline{\includegraphics[scale=0.45]{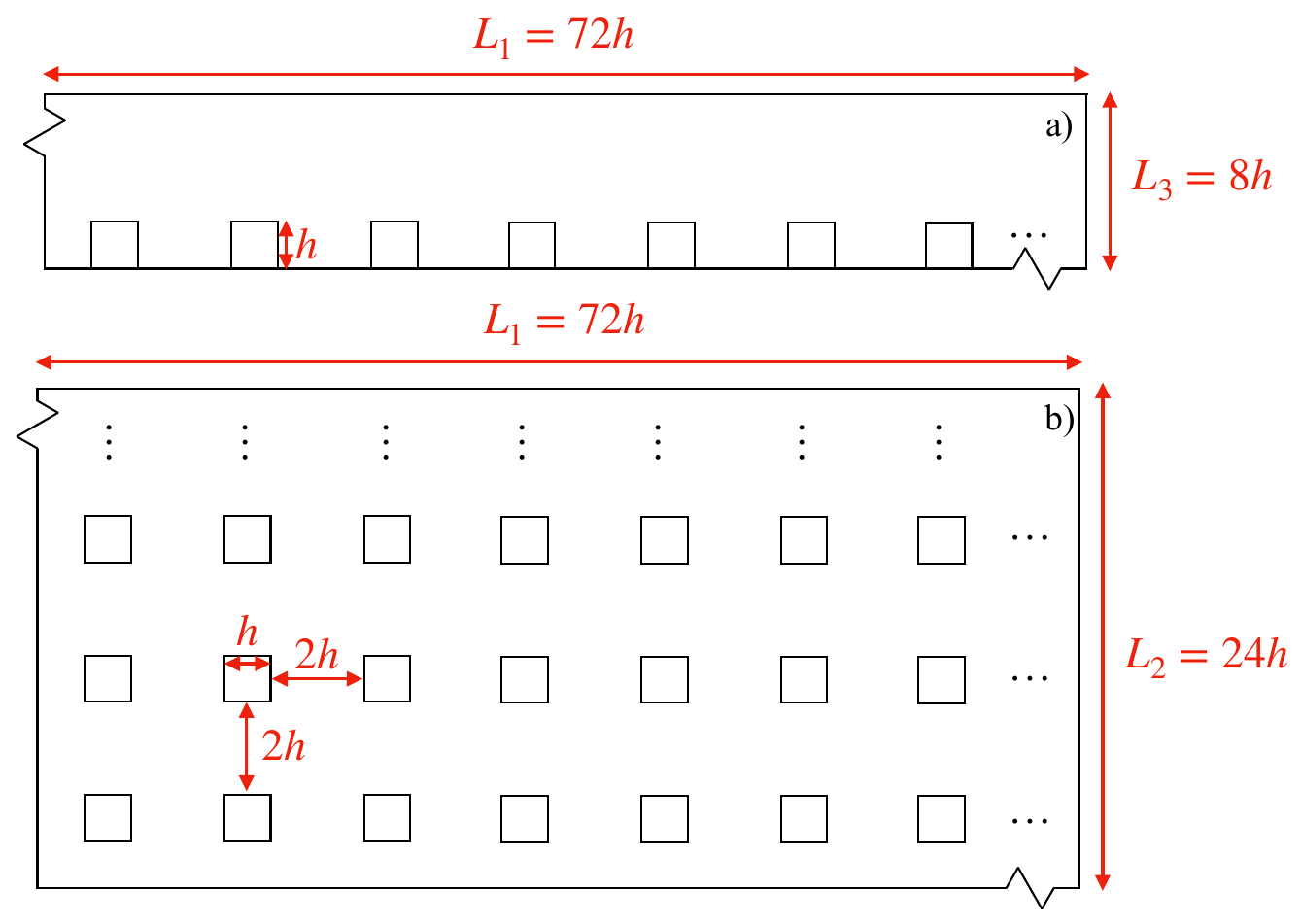}}
  \caption{Side (\textit{a}) and planar view (\textit{b}) of the computational domain.}
\label{fig:flow_config}
\end{figure}

Figure \ref{fig:flow_config} shows a schematic of the computational domain.
The size of the domain is $[0,L_{1}] \times [0,L_{2}] \times [0,L_{3}]$ with $L_{1} = 72h$, $L_{2} = 24h$, and $L_{3} = 8h$, where $h$ is the height of the cuboids. The planar and frontal areas of the cube array are set to $\lambda_p = \lambda_f = 0.\overline{1}$.
An aerodynamic roughness length of $z_0=10^{-4}h$ is prescribed at the cube surfaces and the lower surface via the wall-layer model.
With the chosen value of $z_0$, the SGS pressure drag is a negligible contributor to the overall momentum balance \citep{yang2016recycling}.
The domain is discretized using a uniform Cartesian grid $(N_1, N_2, N_3)= (576, 192, 128)$ where each cube is resolved by $(n_1, n_2, n_3)= (8, 8, 16)$ grid points.
As shown in Appendix~\ref{sec:gird_sensitivity}, this resolution yields flow statistics---up to second-order moments---that are poorly sensitive to grid resolution.

\subsection{Averaging operations}\label{subsec:phase_average}

Given the time-periodic nature of the flow system, phase averaging is the natural approach to evaluate flow statistics in pulsatile flows \citep{scotti2001numerical,bhaganagar2008direct,weng2016numerical,onder2019turbulent}.
Phase averaging can be best understood as a surrogate of Reynolds ensemble averaging for time-periodic flows.
The phase and intrinsic volume average \citep{schmid2019volume} (hereafter referred to as \textit{phase-average} for brevity) of a quantity of interest $\theta$ can be defined as
\begin{equation}
 \langle \theta \rangle(x_3,t)= \frac{1}{N_p}\sum^{N_p}_{n=1} \left( \frac{1}{V_f}\int_{x_3-\delta_3/2}^{x_3+\delta_3/2}\int_0^{L_2} \int_0^{L_1}\theta(x_1,x_2,x_3,t+nT) dx_1 dx_2 dx_3 \right)\ ,\quad 0\leq t \leq T\ ,
	\label{eq:phavg}
\end{equation}
where  $\langle \cdot \rangle$ denotes the phase averaging operation,
$V_f$ is a thin fluid slab of thickness $\delta_3$ in the $x_3$ direction,
$N_p$ denotes the number of the pulsatile cycles over which the averaging operation is performed, and $T=2\pi/\omega$ is the time period of the pulsatile forcing.
A given instantaneous quantity $\theta$ can be decomposed as
\begin{equation}
 \theta (x_1,x_2,x_3,t)= \langle \theta \rangle (x_3,t)+\theta^\prime (x_1,x_2,x_3,t)\ ,
	\label{eq:prime}
\end{equation}
where $(\cdot)^\prime$ denotes a departure of the instantaneous value from the corresponding phase-averaged quantity.
A phase-averaged quantity can be further decomposed into a \textit{longtime average} and an \textit{oscillatory} component with zero mean, i.e.,
\begin{equation}
 \langle \theta \rangle (x_3,t)=\overline{\theta}(x_3)+\widetilde{\theta}(x_3,t)\ .
	\label{eq:bar&tilde}
\end{equation}
This work relies on the \cite{scotti2001numerical} approach to analyze the flow system; in this approach, an oscillatory quantity $\widetilde{\theta}$ is split into two components: one corresponding to the flow oscillation at the forcing frequency (fundamental mode), and one which includes contributions from all of the remaining harmonics, i.e.,
\begin{equation}
\widetilde{\theta}(x_3,t) = A_{\theta}(x_3)\sin \left[ \omega t+\phi_{\theta}(x_3) \right]+ e_\theta(x_3)\ ,
\label{eq:a_phi_def}
\end{equation}
where $A_{\theta}$ and $\phi_\theta$ are the oscillatory amplitude of the fundamental mode and the phase lag with respect to the pulsatile forcing, respectively.
These components are evaluated via minimization of $\|e_\theta\|_2$ at each $x_3$.

\subsection{Dimensional analysis and suite of simulations}\label{subsec:alpha_omega}

Having fixed the domain size, the aerodynamic surface roughness length, and the spatial discretization, the remaining physical parameters governing the problem are:
(i) the oscillation amplitude $\alpha$, defined as the ratio between the oscillation amplitude of $\langle u_1\rangle$ at the top of the domain and the corresponding mean value;
(ii) the forcing frequency $\omega$;
(iii) the friction velocity based on the mean pressure gradient $\overline{u}_{\tau}=\sqrt{f_m L_3}$; and
(iv) the height of roughness elements $h$. 
The latter is a characteristic length scale of the flow in the urban canopy layer (UCL).
System response is studied in time and along the $x_3$ coordinate direction, so
(v) the wall-normal elevation $x_3$ and (vi) time $t$ should also also be included in the parameter set.
Note that the viscosity is not taken into account since the flow is in the fully rough regime.
Choosing $\overline{u}_{\tau}$ and $h$ as repeating parameters, a given normalized longtime $(\overline{Y})$ and phase-average $(\langle Y \rangle)$ quantity of interest can hence be written as

\begin{equation}
\overline{Y}=f(\frac{x_3}{h}, \frac{\omega h}{\overline{u}_\tau} ,\alpha) \ ,  \quad \textrm{and} \quad 
{\langle Y \rangle}=g(\frac{x_3}{h},\omega t,\frac{\omega h}{\overline{u}_\tau},\alpha) \ ,
	\label{eq:timeavg_dimless}
\end{equation}
respectively, where $f$ and $g$ are universal functions. 
(\ref{eq:timeavg_dimless}) show that $\overline{Y}$ and $\langle Y \rangle$ only depend on two dimensionless parameters, namely $\alpha$ and $\omega T_\mathrm{h}$.
$T_\mathrm{h} = h/\overline{u}_{\tau}$ is the turnover time of the largest eddies in the UCL and can be best understood as a characteristic time scale of the flow in the UCL.
$\omega T_\mathrm{h} \sim T_\mathrm{h}/T$ is hence essentially the ratio between the turnover time of the largest eddies in the UCL ($T_\mathrm{h}$) and the pulsation time period ($T$).
Also, note that given their identical mathematical formulation, the normalized $\omega$ can be seen as an equivalent Strouhal number which has been used as a non-dimensional parameter to characterize pulsatile flows in early works on pulsatile flow.
Also, note that $\omega T_\mathrm{h}$ can be seen as an equivalent Strouhal number, which has been used as a non-dimensional parameter to characterize pulsatile flows in earlier studies of this flow system, such as \cite{tardu1994turbulent}, given their identical mathematical formulation.
Four different values of $\omega T_\mathrm{h}$ are considered in this study, namely $w T_{\mathrm{h}}=\{0.05\pi, 0.125\pi, 0.25\pi, 1.25\pi\}$.
For $\overline{u}_{\tau}\approx 0.1 \ \rm{ms^{-1}}$ and $h \approx 10 \ \rm{m}$---common ABL values for these quantities \citep{stull1988introduction}---the considered $\omega T_\mathrm{h}$ set encompasses time scales variability from a few seconds to several hours, which are representative of submeso-scale phenomena \citep{mahrt2009characteristics,hoover2015submeso}.

In this study, we aim to narrow our focus to the \textit{current-dominated} regime, i.e., $0<\alpha<1$.
Larger values of $\alpha$ would lead to a \textit{wave-dominated} regime, which behaves differently.
The specific values $(\alpha = \{ 0.2, 0.4 \})$ have been chosen because they are sufficiently large to lead to an interesting flow response, yet sufficiently far from the wave-dominated regime; this enables us to focus on departures from stationarity induced by flow pulsation in the current-dominated regime.
Due to the lack of a straightforward relation between the oscillatory pressure gradient $(\alpha_p)$ and $\alpha$, $\alpha_p$  has been tuned iteratively to achieve the desired $\alpha$.
As shown in table~\ref{tab:les_runs}, despite the optimization process, there are still discernible discrepancies between the target $\alpha$ and its actual value.
As also shown in previous works \citep{scotti2001numerical, bhaganagar2008direct}, $\alpha$ is highly sensitive to variations in $\alpha_p$, which makes it challenging to obtain the desired $\alpha$ values.  

The suite of simulations and corresponding acronyms used in this study are listed in table~\ref{tab:les_runs}.
A statistically stationary flow case ($\alpha=0$) is also carried out to highlight departures of pulsatile flow cases from the steady state condition.
Simulations with pulsatile forcing are initialized with velocity fields from the stationary flow case; the $\omega  T_\mathrm{h}=\{0.25\pi, 1.25\pi\}$ and $\omega  T_\mathrm{h}=\{0.05\pi, 0.125\pi\}$ cases are then integrated in time over $200 T_{L_3}$ and $400 T_{L_3}$, respectively, where $T_{L_3}= L_3/\overline{u}_\tau$ is the turnover time of the largest eddies in the domain.
This approach yields converged phase-averaged flow statistics.
The size of the time step $\delta t$ is chosen to satisfy the Courant–Friedrichs–Lewy stability condition ${(u \delta t)/\delta_x} \le 0.05$, where $u$ is the maximum velocity magnitude at any given spatial location and time during a run, and $\delta_x$ is the grid stencil in the computational domain. 
Instantaneous three-dimensional snapshots of the velocity and pressure fields are collected every $T/16$ for the $\omega  T_\mathrm{h}=\{0.25\pi, 1.25\pi\}$  cases and every $T/80$ for the $\omega  T_\mathrm{h}=\{0.05\pi, 0.125\pi\}$ cases, after an initial $20 T_{L_3}$ transient period.

\begin{table}
  \begin{center}
\def~{\hphantom{0}}
  \begin{tabular}{lcccc}
        Acronym  & Target $\alpha$ & Actual $\alpha$ & $\alpha_p$  & $\omega T_{\mathrm{h}}$ \\[3pt]
      LL   & 0.2 & 0.17 & 2.4 & 0.05$\pi$  \\
      LM   & 0.2 & 0.16 & 6.0 & 0.125$\pi$  \\
      LH   & 0.2 & 0.16 & 12.0 & 0.25$\pi$  \\
      LVH  & 0.2 & 0.16 & 60.0 & 1.25$\pi$  \\
      HL   & 0.4 & 0.38 & 4.8 & 0.05$\pi$   \\
      HM   & 0.4 & 0.36 & 12.0 & 0.125$\pi$   \\
      HH   & 0.4 & 0.36 & 24.0& 0.25$\pi$   \\
      HVH  & 0.4 & 0.37 & 120.0 & 1.25$\pi$\\
      SS & 0.0 & 0.0 & 0.0 & -  \\
  \end{tabular}
  \caption{List of LES runs. The naming convention for pulsatile flow cases is as follows. The first letter represents the oscillation amplitude: L for $\alpha=0.2$ and H for $\alpha=0.4$.
  The second and third letters denote the forcing frequencies: L for $\omega T_{\mathrm{h}}=0.05\pi$, M for $\omega T_{\mathrm{h}}=0.125\pi$, H for $\omega T_{\mathrm{h}}=0.25\pi$, and VH for $\omega T_{\mathrm{h}}=1.25\pi$. SS denotes the statistically stationary flow case.}
  \label{tab:les_runs}
  \end{center}
\end{table}

\section{Results and discussion}\label{sec:results}

\subsection{Instantaneous velocity field}\label{subsec:inst}

\begin{figure}
  \centerline{\includegraphics[width=\textwidth]{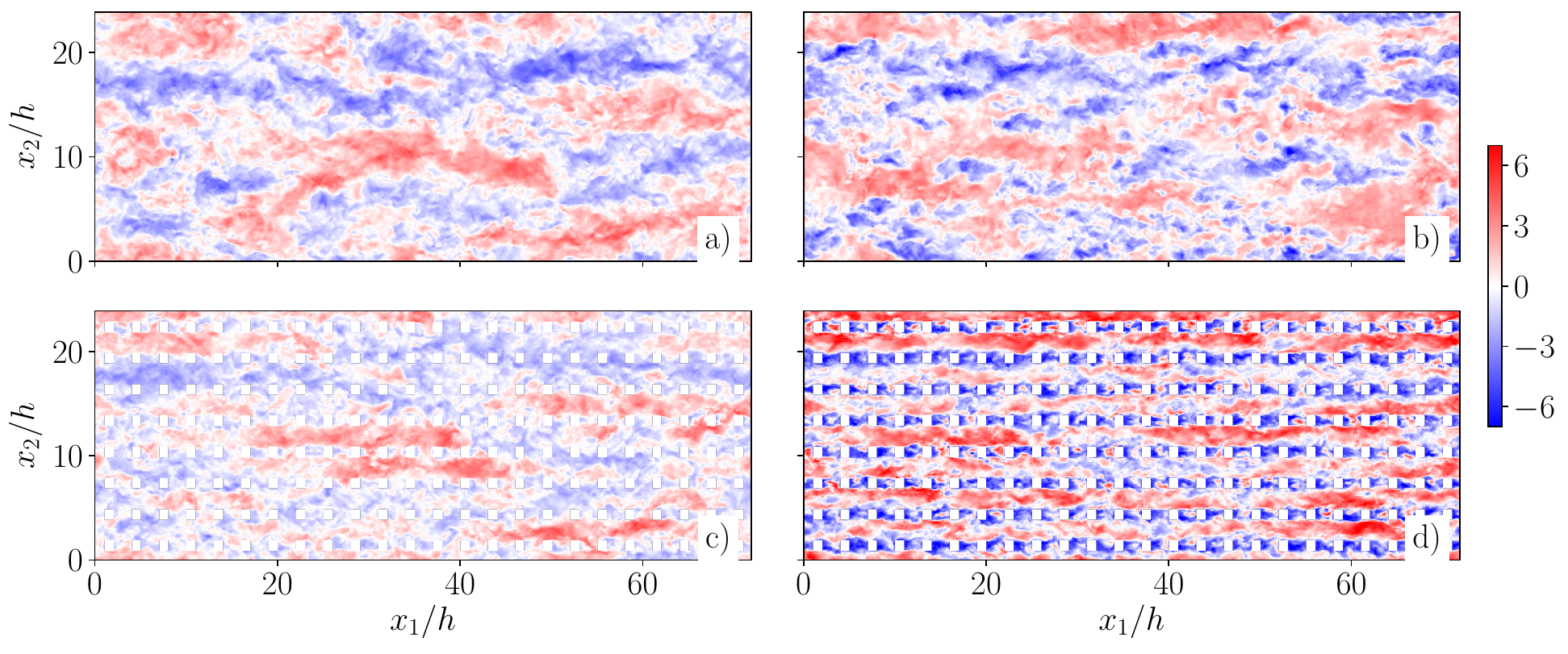}}
  \caption{Instantaneous streamwise fluctuating velocity field at streamwise/cross-stream plane $x_3=2h$ (\textit{a}, \textit{b}) and $x_3=0.75h$ (\textit{c}, \textit{d}) from the HM case. Panels \textit{a} and \textit{c} correspond to $t=0$, whereas panels \textit{b} and \textit{d} to $t=\pi/2$.}
  \label{fig:inst_xy}
\end{figure}

To gain insights into the instantaneous flow field, figure~\ref{fig:inst_xy} displays the streamwise fluctuating velocity within and above the UCL from the HM case at two different $x_3$ planes and phases.
The chosen two phases $t=0$ and $t=\pi/2$ correspond to the end of the deceleration and acceleration period, respectively. 

It is apparent from figure~\ref{fig:inst_xy}(\textit{a}, \textit{b}) that meandering low-momentum ($u^\prime<0$) streaks are flanked by adjacent high-momentum ($u^\prime>0$) ones. 
Comparing these two, turbulence structures at $t=\pi/2$ appear smaller in size in both streamwise and spanwise directions. 
Additionally, within the UCL, apparent vortex shedding occurs on the lee side of the cubes at $t=\pi/2$, while it is less pronounced at $t=0$.
This suggests that flow unsteadiness substantially modifies the flow field during the pulsatile cycle and is expected to impact the flow statistics substantially.

\subsection{Longtime-averaged statistics}\label{subsec:longtime}


\subsubsection{Longtime-averaged velocity profile}\label{subsubsec:longtime_u}

\begin{figure}
  \centerline{\includegraphics[width=\textwidth]{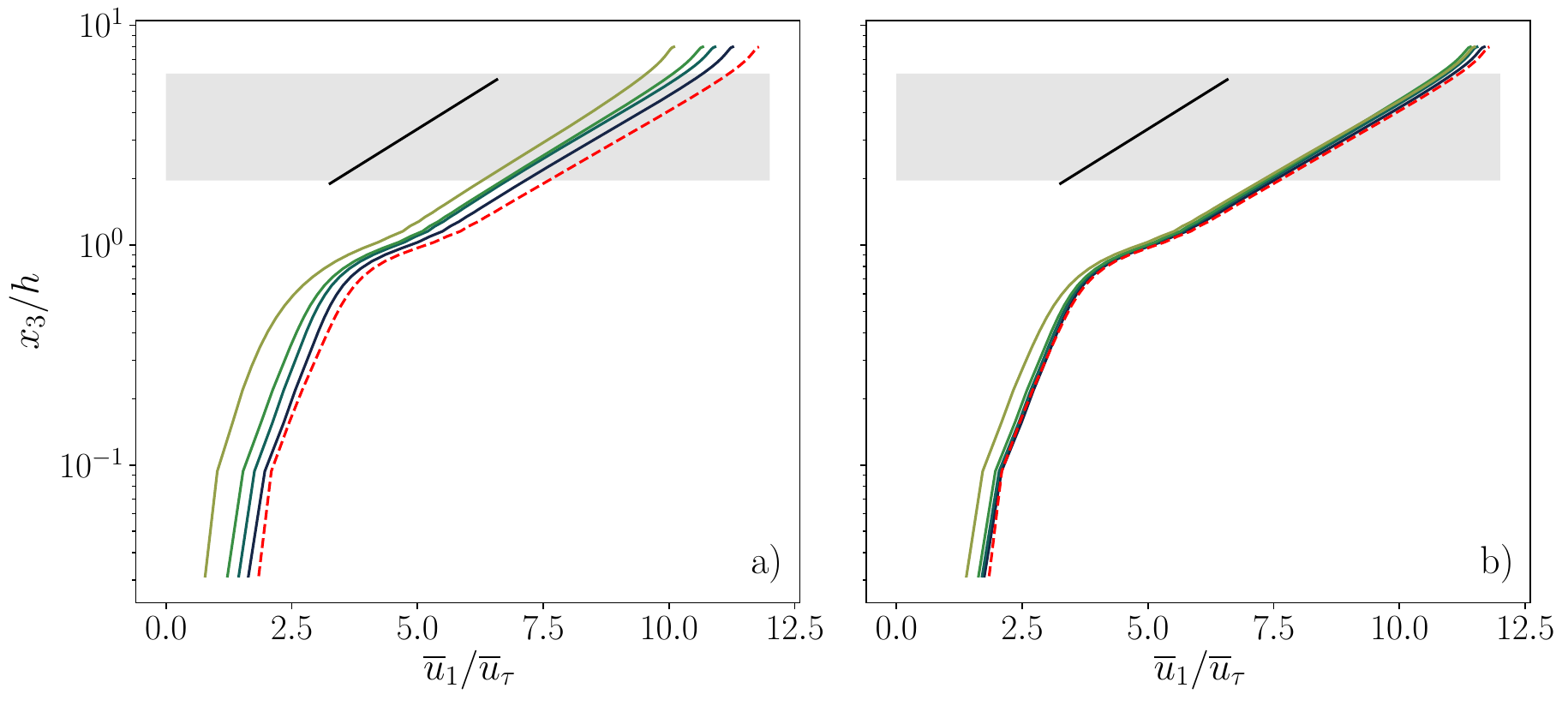}}
  \caption{Wall-normal profiles of longtime-averaged streamwise velocity $\overline{u}_1$ of high-amplitude cases (\textit{a}) and low-amplitude cases (\textit{b}). Line color specifies the forcing frequency: navy blue, $\omega T_{\mathrm{h}}=0.05\pi$; dark green, $\omega T_{\mathrm{h}}=0.125\pi$; light green, $\omega T_{\mathrm{h}}=0.25\pi$; yellow-green, $\omega T_{\mathrm{h}}=1.25\pi$. $\overline{u}_1$ from the SS case, represented by the red dashed line, is included for comparison. Black solid line indicates the slope of $1/\kappa$, with $\kappa=0.4$. The shaded area highlights the fitting region for the estimation of the roughness length scale $z_0$.}
  \label{fig:longtime_uprofile}
\end{figure}

Profiles of the longtime-averaged streamwise velocity are shown in figure \ref{fig:longtime_uprofile}.
Flow unsteadiness leads to a horizontal shift of profiles in the proposed semi-logarithmic plot.
This behavior is distinct from the ``two-log" profile in flow over sand grain roughness \citep{fredsoe1999wave,yang2006velocity,yuan2015experimental}, and also in stark contrast to the one previously observed in current-dominated pulsatile flow over aerodynamically smooth surfaces, where the longtime-averaged field is essentially unaffected by flow unsteadiness \citep{tardu1993wall,Tardu2005,scotti2001numerical,manna2012pulsating,weng2016numerical}.
As also apparent from figure \ref{fig:longtime_uprofile}, departures from the statistically stationary flow profile become more significant for larger values of $\alpha$ and $\omega$.

\begin{figure}
  \centerline{\includegraphics[width=\textwidth]{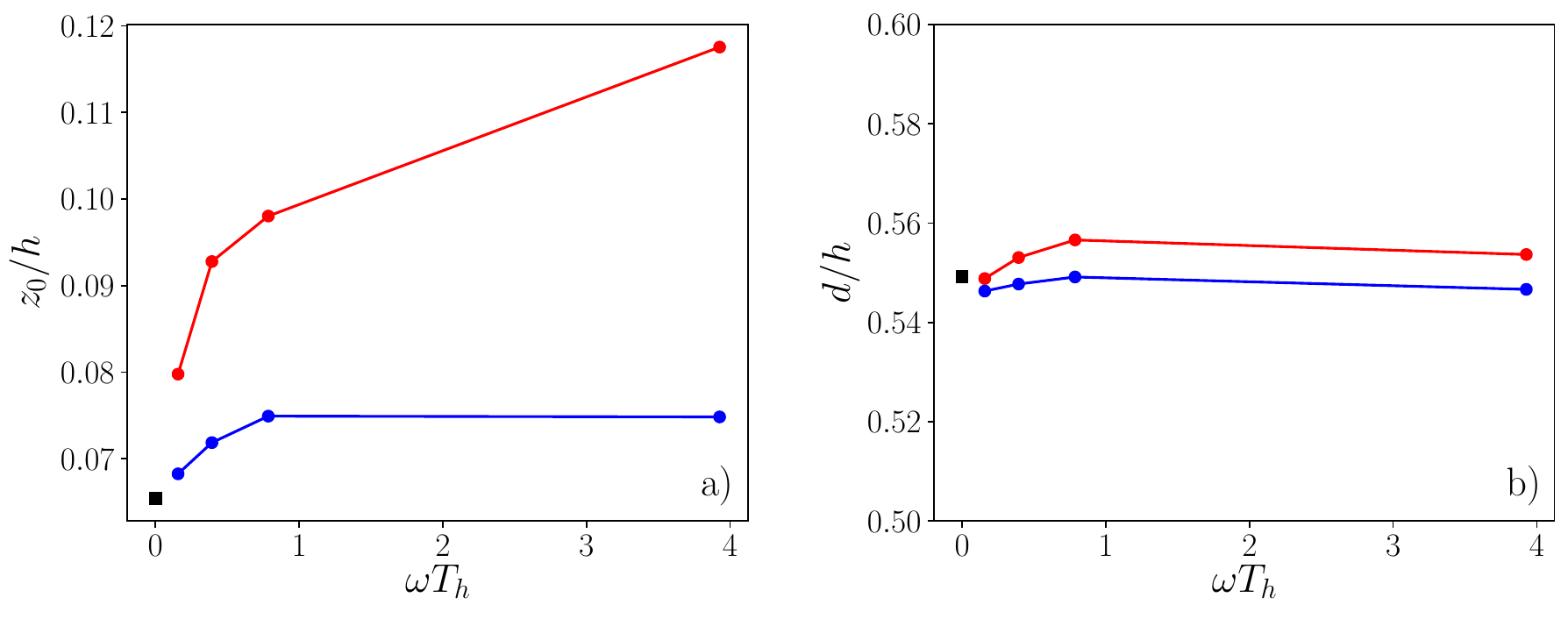}}
  \caption{Normalized aerodynamic roughness length $z_0$ (\textit{a}) and displacement height $d$ (\textit{b}). Different colors correspond to different oscillation amplitudes: blue, $\alpha=0.2$; red, $\alpha=0.4$. The black square symbol denotes the reference SS case.}
\label{fig:z0_d}
\end{figure}

Variations in the aerodynamic roughness length $z_0$ and displacement height $d$ parameters with $\omega T_\mathrm{h}$ are shown in figure \ref{fig:z0_d} for the considered canopy.
These parameters are evaluated via the \cite{macdonald1998improved} approach, where $d$ is the barycenter height of the longtime-averaged pressure drag from the urban canopy, and $z_0$ is determined via curve fitting.
More specifically,
\begin{equation}
 d = \frac{\int^h_0 \overline{D}(x_3)x_3 dx_3 }{\int^h_0  \overline{D}(x_3 )dx_3 }\ ,
	\label{eq:displacement}
\end{equation}
where the wall-normal distribution of the instantaneous canopy pressure drag $D$ is obtained by taking an intrinsic volume average of the pressure gradient, i.e.,
\begin{equation}
D(x_3,t) =\frac{1}{V_f}\int_{x_3-\delta z/2}^{x_3+\delta z/2}\int_0^{L_2} \int_0^{L_1}\frac{1}{\rho}\frac{\partial p}{\partial x} dx_1 dx_2 dx_3\ .
	\label{eq:inst_form}
\end{equation}

Note that, in principle, one should also account for the SGS drag contribution in (\ref{eq:inst_form});
in this work, we omit SGS contributions because they are negligible when compared to the total drag---a direct result of the relatively small aerodynamic roughness length that is prescribed in the wall-layer model (see \S\ref{subsec:num}).

$z_0$ is solved by minimizing the mean square error (MSE) between the longtime-averaged velocity and the law of the wall with $\kappa=0.4$ in the $x_3 \in [2h,6h]$ interval, i.e.,
\begin{equation}
    E=\| \overline{u}_1-\frac{\overline{u}_{\tau}}{\kappa}\log(\frac{x_3-d}{z_0})\|_{2}\ .
    \label{eq:z0}
\end{equation}

The fitting interval is highlighted in figure \ref{fig:longtime_uprofile}.
The estimated $z_0$ was found to be poorly sensitive to variations in the fitting interval within the considered range of values.
\cite{cheng2007flow} argued that MacDonald's method is accurate when surfaces are characterized by a low packing density---a requirement that is indeed satisfied in the considered cases.

As apparent from figure \ref{fig:z0_d}, $d$ is poorly sensitive to variations in both $\alpha$ and $\omega$ (variations across cases are within the $\pm 3\%$ range).
This behavior can be explained by considering that for flows over sharp-edged obstacles, such as the ones considered herein, flow separation patterns are poorly sensitive to variations in $\alpha$ and $\omega$, resulting in a rather constant total volume of wake regions and momentum deficits across cases, and constant longtime-averaged pressure on the surfaces of the cuboids.
The $d$ parameter is here evaluated as integral of the pressure gradient field over the surface area, so the above considerations provide a physical justification for the observed behavior.
Note that this finding might not be generalizable across all possible roughness morphologies.
For instance, \cite{yu2022boundary} showed that separation patterns in pulsatile flows over hemispheres feature a rather strong dependence on $\alpha$ and $\omega$, yielding corresponding strong variations in $d$.

Contrary to $d$, the $z_0$ parameter is strongly impacted by flow unsteadiness, and its value increases with $\alpha$ and $\omega$.
\cite{bhaganagar2008direct} reported a similar upward shift of velocity profile in his simulations of pulsatile flow over transitionally rough surfaces at a low Reynolds number.
She attributed the increase in $z_0$ to the resonance between the unsteady forcing and the vortices shed by roughness elements, which is induced when the forcing frequency approaches that of the vortex shedding.
However, such an argument does not apply to the cases under investigation, since we observed no spurious peaks in the temporal streamwise velocity spectrum.
Rather, the increase in $z_0$ stems from the quadratic relation between the phase-averaged canopy drag and velocity, as elaborated below.

\subsubsection{Phase-averaged drag-velocity relation}\label{subsubsec:du_relation}

\begin{figure}
  \centerline{\includegraphics[width=\textwidth]{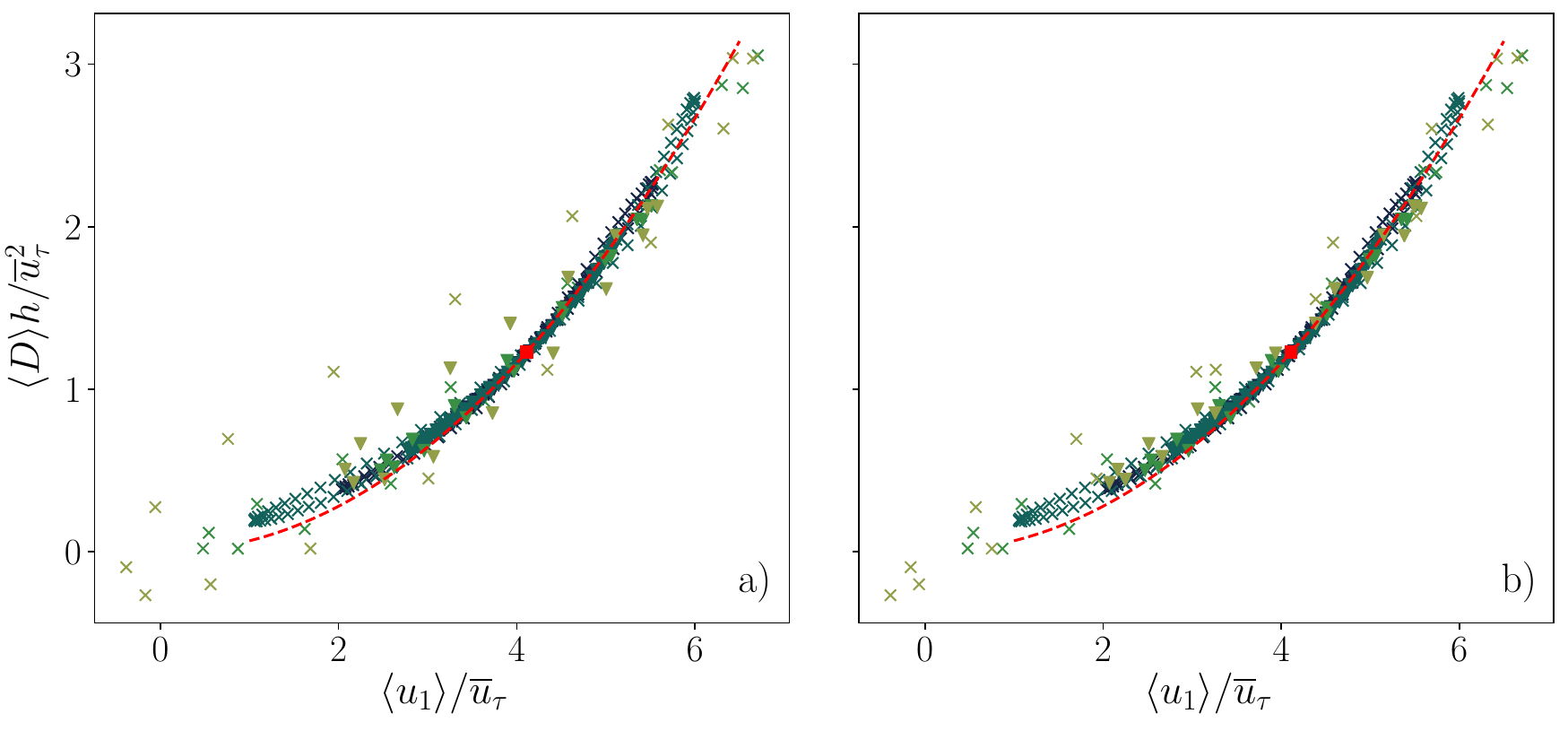}}
  \caption{Phase-averaged canopy drag $\langle D\rangle$ at $x_3/h\approx 0.8$ as a function of the local phase-averaged velocity $\langle u_1\rangle$ (\textit{a}) and the same $\langle D\rangle-\langle u_1 \rangle$ plot but with the time lag between $\langle D\rangle$ and $\langle u_1 \rangle$ \textit{removed} (\textit{b}). Different colors are used to denote different forcing frequencies: navy blue, $\omega T_{\mathrm{h}}=0.05\pi$; dark green, $\omega T_{\mathrm{h}}=0.125\pi$; light green, $\omega T_{\mathrm{h}}=0.25\pi$; yellow-green, $\omega T_{\mathrm{h}}=1.25\pi$. Different symbols are used to distinguish between oscillation amplitudes: triangle, $\alpha=0.2$; cross, $\alpha=0.4$. Red square represents the SS case. Red dashed line represents $\langle D\rangle=C_d\lambda_f \langle u_1 \rangle \left | \langle u_1 \rangle \right |$ with $C_d$ obtained from the SS case.}
\label{fig:d_u}
\end{figure}

\begin{figure}
  \centerline{\includegraphics[width=\textwidth]{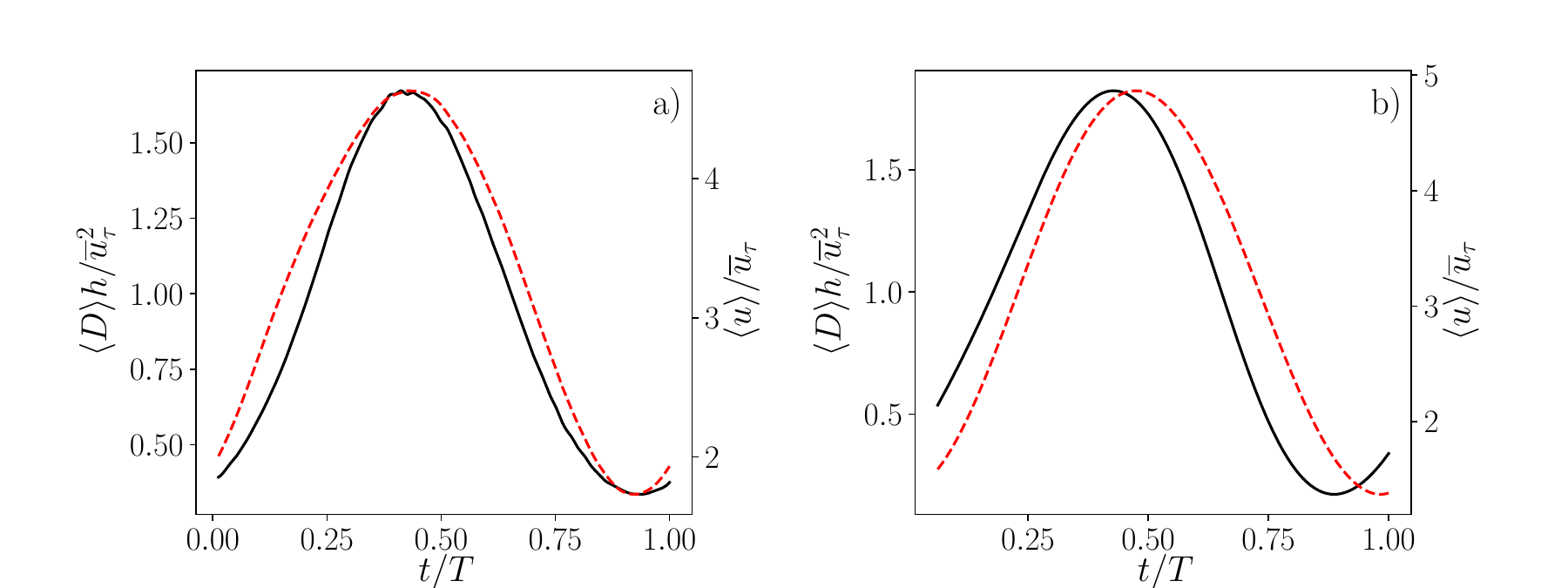}}
  \caption{Time evolution of $\langle D\rangle $ (black solid lines) and $\langle u_1 \rangle$ (red dashed lines) at $x_3/h\approx 0.8$ from the HL (\textit{a}) and LVH (\textit{b}) cases. The acronyms of LES runs are defined in table~\ref{tab:les_runs}.}
  \label{fig:d_u_phaselag}
\end{figure}

The phase-averaged canopy drag $\langle D\rangle$ and the local phase-averaged velocity $\langle u_1 \rangle$ at $x_3/h\approx 0.8$ are shown in figure \ref{fig:d_u}, and are representative of corresponding quantities at different heights within the UCL.
Results from cases with three lower frequencies and that from the SS case cluster along a single curve, highlighting the presence of a frequency-independent one-to-one mapping between $\langle D\rangle$ and $\langle u_1 \rangle$.
As apparent from figure \ref{fig:d_u_phaselag}(\textit{a}), at these three forcing frequencies, the interaction between the wind and the canopy layer is in a state of quasi-equilibrium, i.e., $\langle D\rangle$ is in phase with $\langle u_1 \rangle$.
Moreover, the shape of the aforementioned curve generally resembles the well-known quadratic drag law, which is routinely used to parameterize the surface drag in reduced order models for stationary flow over plant and urban canopies \citep{lettau1969note,raupach1992drag,katul2004one,poggi2004momentum,macdonald1998improved,coceal2004canopy}.
This finding comes as no surprise, given the quasi-equilibrium state of the $\langle D\rangle-\langle u_1 \rangle$ relation for the three lower forcing frequencies.

On the other hand, as shown in figure \ref{fig:d_u_phaselag}(\textit{b}), results from the highest-frequency cases (LVH and HVH) exhibit an orbital pattern, which stems from the time lag between $\langle D\rangle$ and $\langle u_1 \rangle$.
Artificially removing this time lag indeed yields a better clustering of data points from the highest-frequency cases along the quadratic drag law (see figure \ref{fig:d_u}\textit{b}).

These findings suggest the following parameterization for $\langle D\rangle$:
\begin{equation}
    \langle D \rangle(x_3,t)= C_d(x_3) \lambda_f \langle u_1 \rangle \left | \langle u_1 \rangle \right |(x_3,t+\Delta t)\ ,
    \label{eq:d_u_relationship}
\end{equation}
where $C_d$ is a sectional drag coefficient that is constant in time and does not depend on $\alpha$ nor $\omega$, and $\Delta t$ accounts for the time lag between $\langle D\rangle$ and $\langle u_1 \rangle$, which instead does indeed depend on $\alpha$ and $\omega$.
Note that, throughout the considered cases, the wall-normal-averaged drag coefficient $\int^{h}_0 C_d dx_3/h\approx0.9$---a value that is similar to those previously reported ones for stationary flow over cube arrays \citep{coceal2004canopy} (note that the exact value depends on the formula used to define $C_d$).

\cite{morison1950force} developed a semi-empirical model relating the phase-averaged drag generated by obstacles in an oscillatory boundary layer to a given phase-averaged velocity---a model that has been extensively used in the ocean engineering community to evaluate drag from surface-mounted obstacles \citep{lowe2005oscillatory,yu2018interaction,yu2022boundary}.
The Morison model assumes that the total force applied to the fluid by obstacles consists of a quadratic drag term and an inertial term; the latter accounts for the added mass effect and the Froude-Krylov force arising as a direct consequence of the unsteady pressure field.
One might argue that the Morison model could also be used to evaluate surface drag as a function or the phase-averaged velocity at a given $x_3$ for the cases under consideration, but unfortunately, this is not the case.
As shown in \cite{patel2013compliant}, the Morison model provides relatively accurate evaluations of obstacle drag when the phased-averaged acceleration at different $x_3$ are in phase.
This is not the case in this study, where important phase lags between phase-averaged accelerations at different wall-normal locations substantially degrade the accuracy of such a model.
This behavior can be easily inferred from figure \ref{fig:contour_u}.

In the following, we will make use of (\ref{eq:d_u_relationship}) to derive an alternative phenomenological surface-drag model for the considered flow system.

\subsubsection{Mapping roughness length variability to longtime-averaged flow statistics}\label{subsubsec:z0model}

$z_0$ and $d$ are input parameters of surface flux parameterizations that are routinely used in numerical weather prediction, climate projection, and pollutant dispersion models \citep[see, e.g.][]{Shamarock2008, Benjamin2016}. 
These models are typically based on Reynolds-averaged Navier-Stokes closures, and feature time steps that can go from one hour up to several days. 
When departures from stationarity occur at a time scale that is much smaller than the time step of the model, model predictions are essentially longtime-averaged quantities, and the validity of surface flux parameterizations based on flow homogeneity and stationarity assumptions may break down.  
As a first step towards addressing this problem, this section proposes a phenomenological model relating $z_0$ to longtime-averaged pulsatile-flow statistics. 

The longtime-averaged friction velocity can be written as
\begin{equation}
    \overline{u}^2_{\tau}=\int^{h}_0 \overline{D}  dx_3=\int^{h}_0  C_d \lambda_f \overline{\langle u_1 \rangle \left | \langle u_1 \rangle \right |}dx_3\ ,
    \label{eq:total_d_overline}
\end{equation}
where $\overline{D}$ is the longtime-averaged surface drag.
Note that the wall-normal structure of $C_d$ is approximately constant in the UCL (not shown here), except in the vicinity of the surface, where local contributions to the overall drag are however minimal due to the small value of $\langle u_1 \rangle$.
Thus, it is reasonable to assume $C_d$ is constant along the wall-normal direction.
Also, depending on $\alpha$ and $\omega$, the flow within the canopy might undergo a local reversal in the phase-averaged sense, meaning that $\langle u_1 \rangle < 0$ at selected $x_3$ locations.
Assuming that there is no flow reversal within the UCL, i.e., $\langle u_1 \rangle \ge 0$ in $z \le h$, (\ref{eq:total_d_overline}) can be written as
\begin{equation}
     \overline{u}_{\tau}^2= C_d \lambda_f \left( \int^{h}_{0}{\overline{u}_1^2}{dx_3} + \int^{h}_{0}{\overline{\widetilde{u}_1^2}}{dx_3} \right) \ ,
    \label{eq:total_d_noflowreversal}
\end{equation}
where the second term on the right-hand side of (\ref{eq:total_d_noflowreversal}) is identically zero for the SS case.
(\ref{eq:total_d_noflowreversal}) essentially states that an unsteady canopy layer requires a lower longtime-averaged wind speed to generate the same drag of a steady canopy layer since quadratic drag contributions are generated by flow unsteadiness (the second term on the right-hand side of (\ref{eq:total_d_noflowreversal})).
Note that $\sqrt{\int^{h}_0 (\cdot)^2 dx_3/h }$ is an averaging operation over the UCL based on the $L_2$ norm.
Rearranging terms in (\ref{eq:total_d_noflowreversal}) leads to
\begin{equation}
    \overline{u}_{1,\mathrm{avg}} = \sqrt{\frac{\overline{u}^2_{\tau}}{C_d \lambda_f h} -\frac{1}{h}\int^{h}_0 \overline{\widetilde{u}_1^2} dx_3}\ ,    
    \label{eq:u_overline_puls}
\end{equation}
and 
\begin{equation}
    \overline{u}_{1,\mathrm{avg}}^{\mathrm{SS}} = \sqrt{\frac{\overline{u}^2_{\tau}}{C_d \lambda_f h}} \ ,    
    \label{eq:u_overline_nonpuls}
\end{equation}
For the pulsatile cases and the SS case, respectively. 
Here $(\cdot)_{\mathrm{avg}}$ denotes the canopy-averaged quantity, and $(\cdot)^{\mathrm{SS}}$ represents a quantity pertaining to the SS case. 

\begin{figure}
  \centerline{\includegraphics[width=\textwidth]{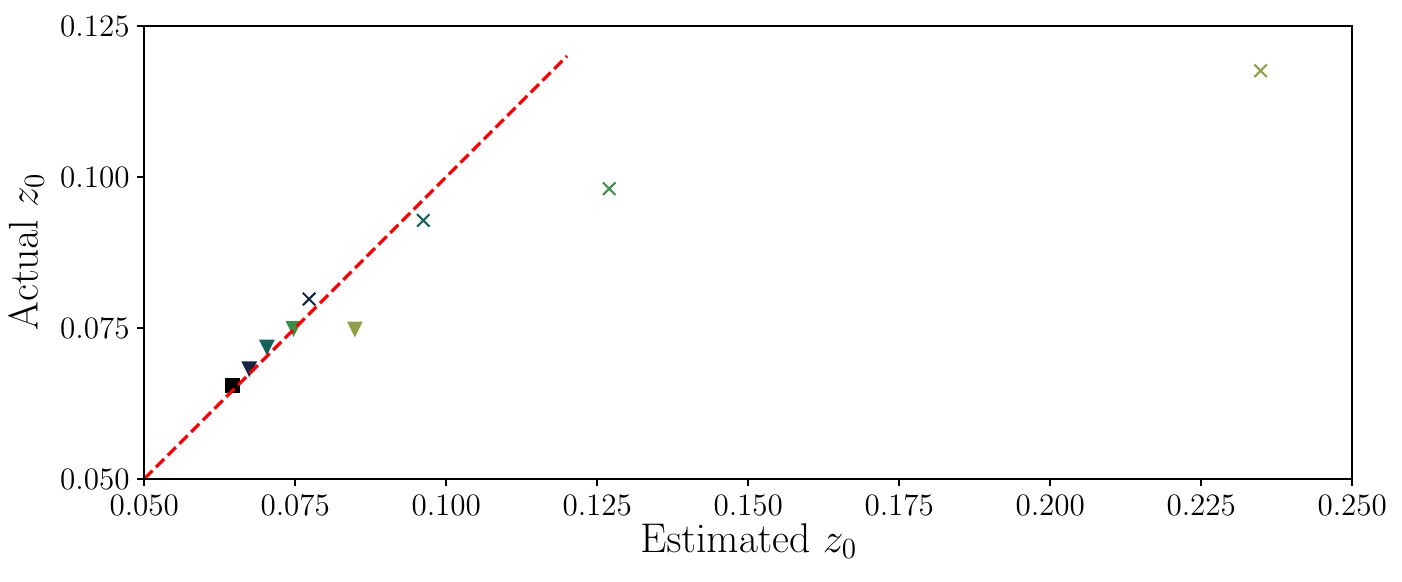}}
  \caption{Comparison between $z_0$ estimated via (\ref{eq:z0_exp}) and $z_0$ from LES. Symbols and colors correspond to those used in figure \ref{fig:d_u}.}
\label{fig:est_z0}
\end{figure}

As discussed in \S\ref{subsubsec:longtime_u}, flow unsteadiness yields a shift of the $\overline{u}_1$ profile with negligible variations in the $d$ parameters when compared to the stationary flow with the same $\overline{u}_\tau$. 
In terms of the law-of-the-wall, this behavior can be described as a variation in $z_0$, i.e., 
\begin{equation}
    \overline{u}_1^{\mathrm{SS}}-\overline{u}_1=\frac{\overline{u}_\tau}{\kappa}\log \left( \frac{x_3-d}{z_0^{\mathrm{SS}}}\right)-\frac{\overline{u}_\tau}{\kappa}\log \left( \frac{x_3-d}{z_0}\right) \ ,
    \label{eq:u_shift}
\end{equation}
where (\ref{eq:u_shift}) is valid for any $x_3$ in the logarithmic region.
The shift in the velocity profile is approximately constant for $x_3 \in [0,L_3]$, so one can write
\begin{equation}
\overline{u}_1^{\mathrm{SS}}-\overline{u}_1 \approx \overline{u}_{1,\mathrm{avg}}^{\mathrm{SS}}-\overline{u}_{1,\mathrm{avg}} \ ,
\label{eq:u_shift_approx}
\end{equation}
and substituting (\ref{eq:u_overline_puls})-(\ref{eq:u_shift}) into (\ref{eq:u_shift_approx}) finally yields
\begin{equation}
    z_0=z_0^{\mathrm{SS}} \exp \left[ \kappa \left(\sqrt{\frac{1}{C_d \lambda_f h}}-\sqrt{\frac{1}{C_d \lambda_f h}-\frac{\int^{h}_0 \overline{\widetilde{u}_1^2}dx_3/h}{\overline{u}^2_{\tau}}}\right)\right]\ .
    \label{eq:z0_exp}
\end{equation}
(\ref{eq:z0_exp}) is a diagnostic model relating variations in the $z_0$ parameter to the UCL phase-averaged velocity variance---a longtime-averaged quantity. 
$z_0$ estimates from (\ref{eq:z0_exp}) are compared against LES results in figure \ref{fig:est_z0}, using $C_d = 0.9$.
It is apparent that the proposed model is able to accurately evaluate $z_0$ for most of the considered cases. 
For the LVH, HH, and HHVW runs, $z_0$ is overestimated by the model; these departures are attributed to the presence of flow reversal in the UCL, which contradicts the model assumptions.
(\ref{eq:z0_exp}) highlights that, in the absence of flow reversal, $z_0$ can be described as a monotonically increasing function of the $\widetilde{u}_1$ variance in the UCL.
As explained at the beginning of this section, this finding is important from a flow modeling perspective, because it relates a longtime-averaged flow statistic to the $z_0$ parameter.
Note that $z_0^{\mathrm{SS}}$ can be accurately evaluated using any existing parameterization for stationary ABL flow over aerodynamically rough surfaces, including the \cite{lettau1969note}, \cite{raupach1992drag}, and \cite{macdonald1998improved} models.
Further, $C_d = 0.9$ (see discussion in \S\ref{subsubsec:du_relation}) and $\lambda_f$ and $h$ are morphological parameters that are in general a-priori available.
In weather forecasting and climate models, $\widetilde{u}_1$ is an SGS quantity and would hence have to be parameterized as a function of longtime-averaged statistics that the model computes or from available in-situ or remote sensing measurements.

\subsubsection{Longtime-averaged resolved Reynolds stress}\label{subsubsec:longtime_reystress}

This section shifts the attention to longtime-averaged resolved Reynolds stresses.
For all of the considered cases, contributions from SGS stresses account for $<1\%$ of the total phase-averaged Reynolds stresses and are hence not discussed.

\begin{figure}
  \centerline{\includegraphics[width=\textwidth]{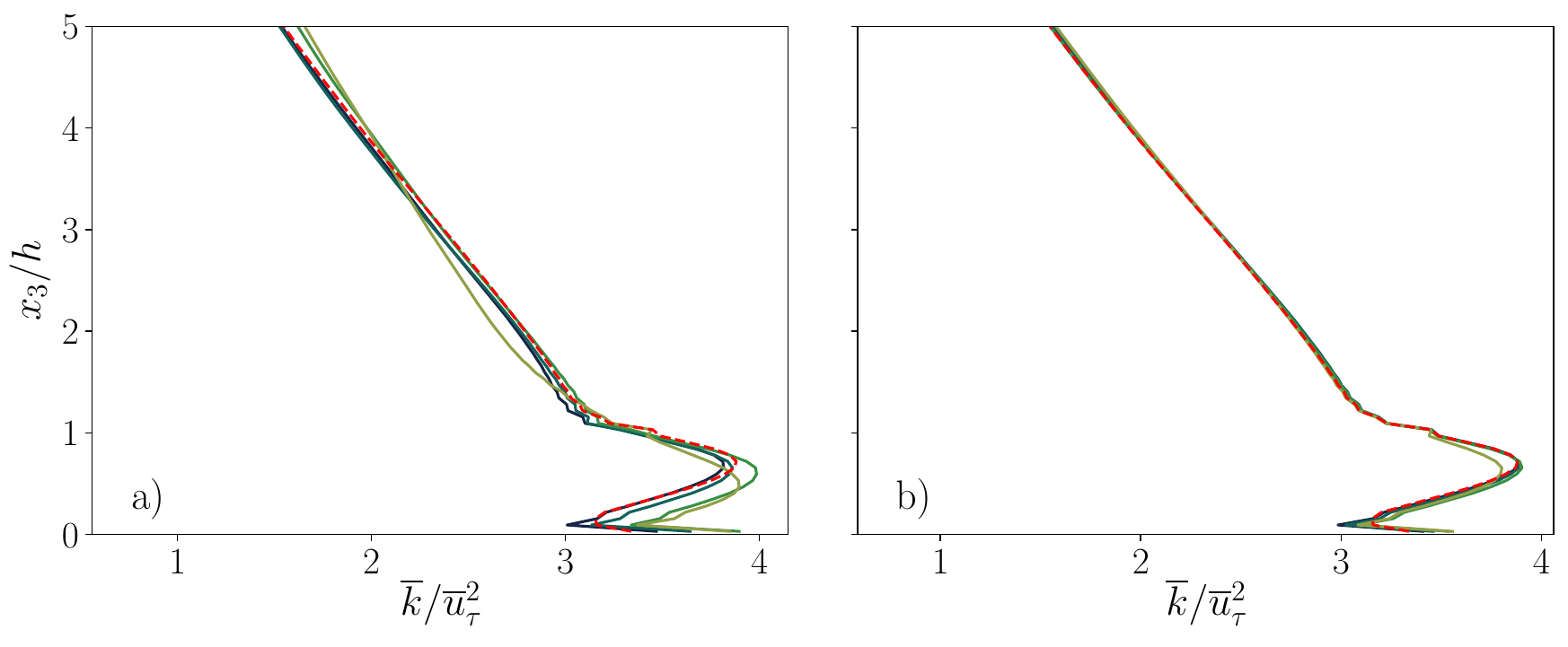}}
  \caption{Longtime-averaged resolved turbulent kinetic energy $\overline{k}=(\overline{u_1^\prime u_1^\prime}+\overline{u_2^\prime u_2^\prime}+\overline{u_3^\prime u_3^\prime})/2$ from high-amplitude cases (\textit{a}) and low-amplitude cases (\textit{b}). Line colors correspond to those used in figure \ref{fig:longtime_uprofile}.}
\label{fig:k_longtime}
\end{figure}

Throughout the boundary layer, $\overline{u_1^\prime u_3^\prime}$ profiles are indistinguishable from the SS one (not shown), indicating a weak dependence of such a quantity on $\alpha$ and $\omega$. 
Above the UCL, the divergence of $\overline{u_1^\prime u_3^\prime}$ balances the longtime-averaged driving pressure gradient $f_m$, i.e., 
\begin{equation}
-\frac{\partial \overline{u_1^\prime u_3^\prime}}{\partial x_3}+f_m=0 \ .
	\label{eq:longtime_mom_bdg_above_ucl}
\end{equation}
Since $f_m$ does not vary across the considered cases and $\overline{u_1^\prime u_3^\prime}(L_3) = 0$, a collapse of $\overline{u_1^\prime u_3^\prime}$ profiles in this region was to be expected from the mathematical structure of the governing equations.  
Within the UCL, the longtime-averaged momentum budget reads
\begin{equation}
-\frac{\partial \overline{u_1^\prime u_3^\prime}}{\partial x_3}+f_m-\overline{D}=0\ .
	\label{eq:longtime_mom_bdg_in_ucl}
\end{equation}
Given that $\overline{u_1^\prime u_3^\prime}$ profiles collapse in this region, $\overline{D}$ is also expected to feature a weak dependence on $\alpha$ and $\omega$, which in turn explains the weak variations in the $d$ parameter that were observed in \S\ref{subsubsec:longtime_u}.

\begin{figure}
  \centerline{\includegraphics[width=\textwidth]{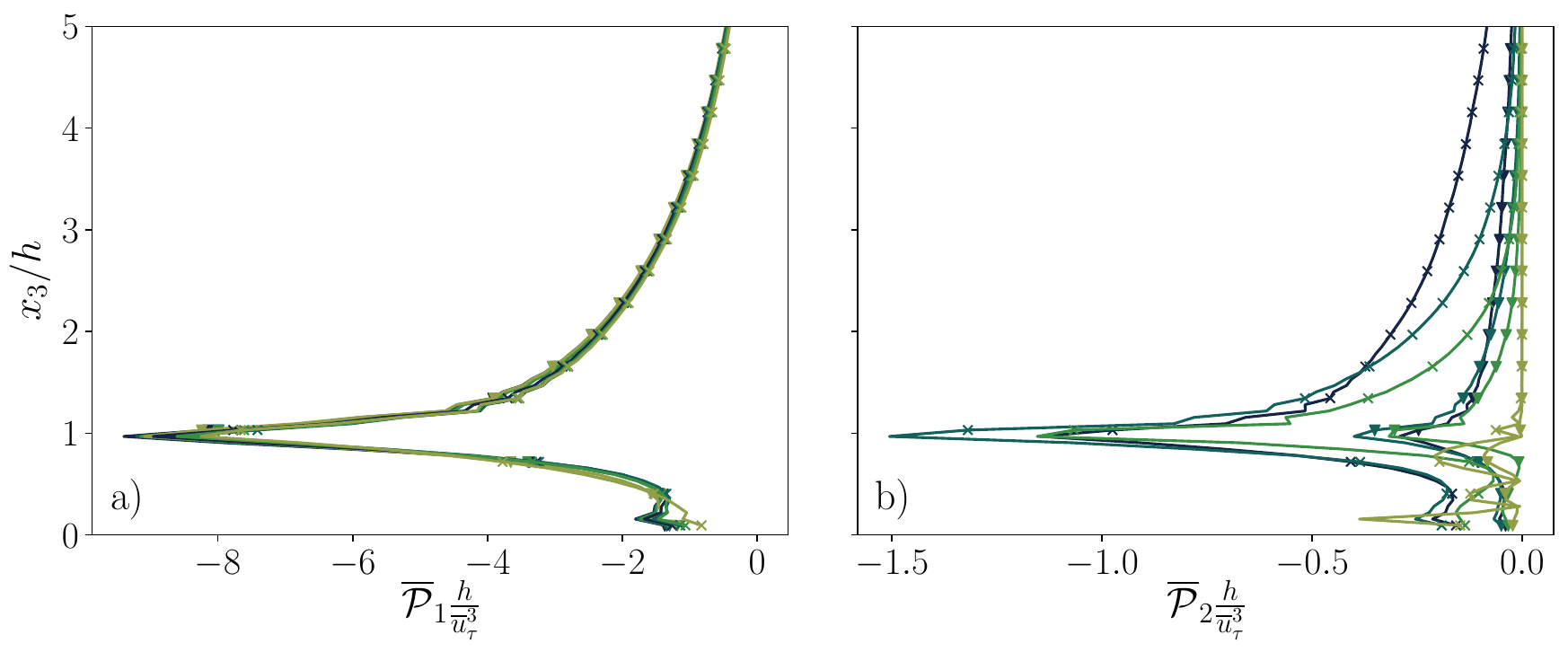}}
  \caption{Normalized shear production terms of $\overline{k}$: $\overline{\mathcal{P}}_1$ (\textit{a}) and $\overline{\mathcal{P}}_2$ (\textit{b}). Symbols and line colors correspond to those used in figure \ref{fig:d_u}.}
\label{fig:longtime_var_prod}
\end{figure}

Figure \ref{fig:k_longtime} depicts wall-normal profiles of the longtime-averaged resolved turbulent kinetic energy, which is defined as
\begin{equation}
\overline{k}=(\overline{u_1^\prime u_1^\prime}+\overline{u_2^\prime u_2^\prime}+\overline{u_3^\prime u_3^\prime})/2\ .
	\label{eq:tke_longtime}
\end{equation}
Such a quantity features a relatively rapid increase in the UCL, which as discussed in Schmid et al. \citep{schmid2019volume}, is due to dispersive contributions caused by the canopy geometry. wh
Also note that the rapid decrease in the near-surface region signals the presence of variations over scales of variability smaller than the vertical grid stencil \citep{coceal2007structure}.  
Flow unsteadiness has a relatively more important impact on such a quantity, especially for flow in the UCL and for cases with a high oscillation amplitude. 
An increase in $\alpha$ generally results in more pronounced departures of $\overline{k}$ profiles from the SS one.
This behavior can be best explained by considering the shear production terms in the budget equation for $\overline{k}$, i.e., 
\begin{equation}
\overline{\mathcal{P}}=-\overline{\langle u_1^\prime u_3^\prime \rangle \frac{\partial \langle u_1 \rangle }{\partial x_3}}=\underbrace{-\overline{ u_1^\prime u_3^\prime} \frac{\partial \overline{u}_1 }{\partial x_3}} _{\overline{\mathcal{P}}_1}\underbrace{-\overline{\widetilde{ u_1^\prime u_3^\prime } \frac{\partial \widetilde{ u }_1 }{\partial x_3}}} _{\overline{\mathcal{P}}_2}\ ,
    \label{eq:longtime_tke_prod}
\end{equation}
where ${\overline{\mathcal{P}}_1}$ represents the work done by $\overline{ u_1^\prime u_3^\prime}$ onto the longtime-averaged flow field, and ${\overline{\mathcal{P}}_2}$ is the longtime average of the work done by the oscillatory shear stress $\widetilde{ u_1^\prime u_3^\prime }$ onto the oscillatory flow field.
Figure \ref{fig:longtime_var_prod}(\textit{a}) shows that ${\overline{\mathcal{P}}_1}$ is poorly sensitive to variations in $\alpha$ and $\omega$.
This behavior stems from the constancy of $\overline{ u_1^\prime u_3^\prime}$ and ${\partial \overline{u}_1 }/{\partial x_3}$ across cases (the latter can be inferred from the systematic shift of $\overline{u}_1$ profiles in figure \ref{fig:longtime_uprofile}).
Conversely, ${\overline{\mathcal{P}}_2}$ from high-amplitude cases are generally larger than those from low-amplitude ones, mainly due to the higher $\widetilde{ u_1^\prime u_3^\prime }$ and $\widetilde{ u}_1$ values.
Discrepancies in ${\overline{\mathcal{P}}_2}$ among high-amplitude cases are larger than those among low-amplitude ones, which ultimately yields the observed variability in $\overline{k}$.

\begin{figure}
  \centerline{\includegraphics[width=\textwidth]{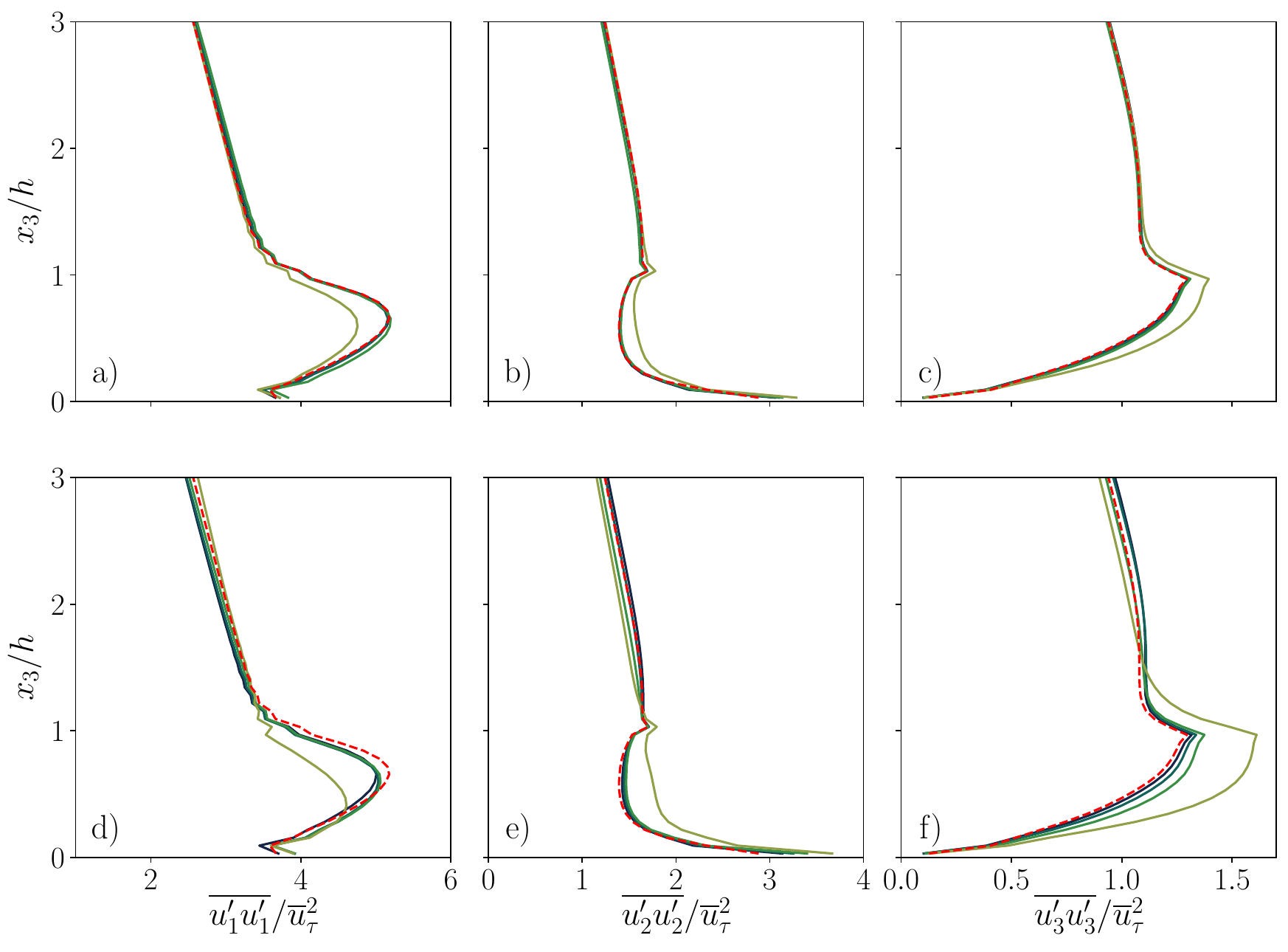}}
  \caption{Longtime-averaged resolved normal Reynolds stresses $\overline{u_1^\prime u_1^\prime}$, $\overline{u_2^\prime u_2^\prime}$, and $\overline{u_3^\prime u_3^\prime}$ from low-amplitude cases (\textit{a}, \textit{b}, \textit{c}) and high-amplitude cases (\textit{d}, \textit{e}, \textit{f}). Line colors correspond to those used in figure \ref{fig:longtime_uprofile}.}
\label{fig:var_longtime}
\end{figure}

\begin{figure}
  \centerline{\includegraphics[width=\textwidth]{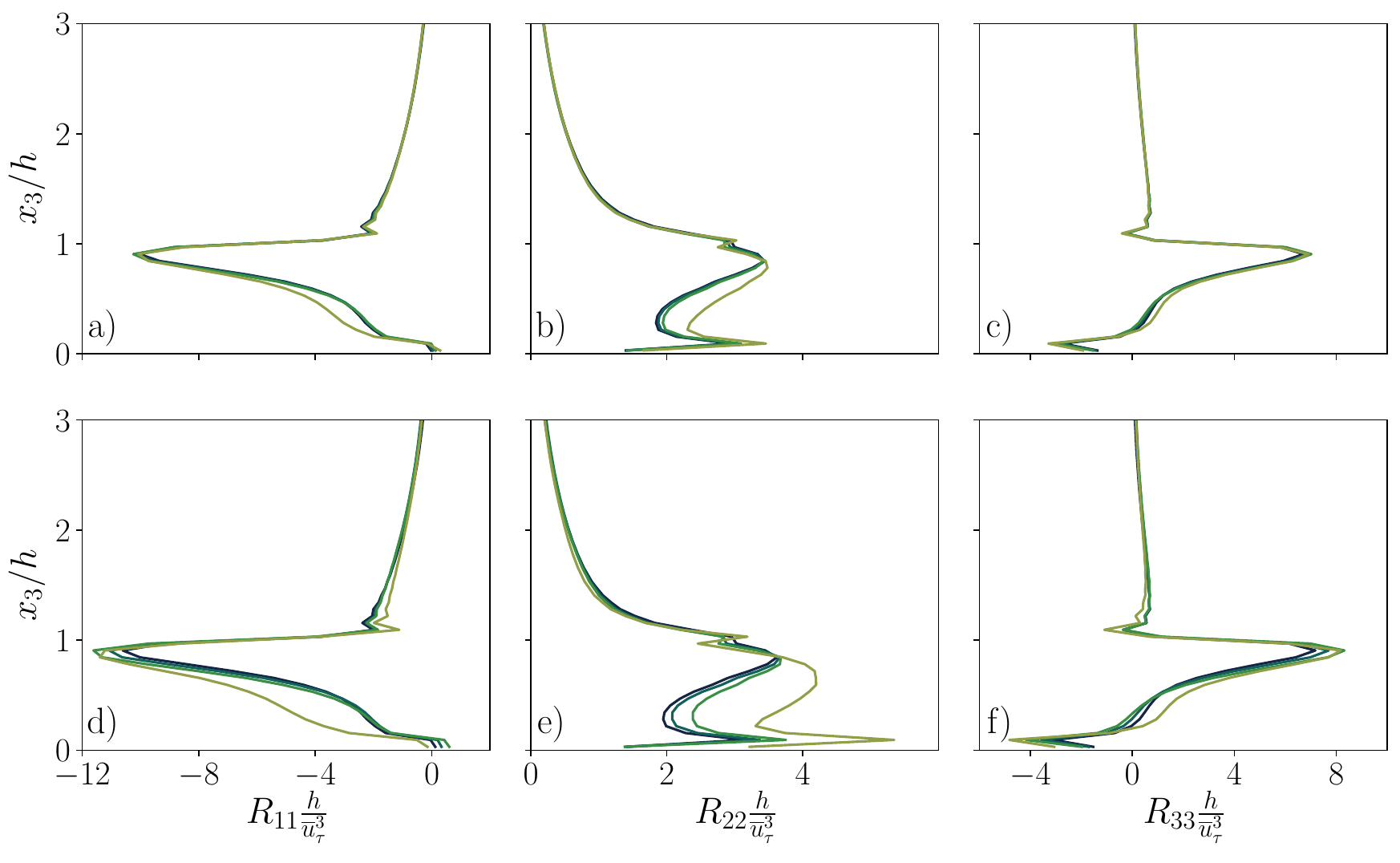}}
  \caption{Pressure redistribution terms $R_{ii}$. (\textit{a}, \textit{b}, \textit{c}): low-amplitude cases; (\textit{d}, \textit{e}, \textit{f}): high-amplitude cases. Line colors correspond to those used in figure \ref{fig:longtime_uprofile}.}
\label{fig:longtime_var_ps}
\end{figure}

Further insight into the problem can be gained by looking at the normal components of the longtime-averaged resolved Reynolds stress tensor, which are shown in figure \ref{fig:var_longtime}.
In this case, it is apparent that increases in the oscillation frequency lead to a decrease in $\overline{u_1^\prime u_1^\prime}$ and an increase in $\overline{u_2^\prime u_2^\prime}$ and $\overline{u_3^\prime u_3^\prime}$ within the UCL---a behavior that is especially apparent for the high-amplitude cases (figure \ref{fig:var_longtime}(e,d,f)).
These trends can be best understood by examining the pressure-strain terms from the budget equations of the longtime-averaged resolved Reynolds stresses, i.e., 
\begin{equation}
R_{ij} = \overline{\frac{p^\prime}{\rho} (\frac{\partial u^\prime_j}{\partial x_i}+\frac{\partial u^\prime_i}{\partial x_j})}
    \label{eq:pressure_reditribution}
\end{equation}

$R_{ii}$ are responsible for redistributing kinetic energy among the longtime-averaged normal Reynolds stresses \citep{Pope2000}, and are shown in figure  \ref{fig:longtime_var_ps}. 
With the exception of the very near surface region ($x_3 \lessapprox 0.2$) where no clear trend can be observed, increases in $\omega$ and $\alpha$ yield a decrease in $R_{11}$ and increase in $R_{22}$ and $R_{33}$, which justify the observed isotropization of turbulence in the UCL.

\subsection{Oscillatory fields}\label{sec:osc_fields}

This section shifts the focus to the time evolution of velocity and resolved Reynolds stresses during the pulsatile cycle.

\subsubsection{Oscillation amplitude impacts on the oscillatory fields}\label{sec:osc_fields_amp}

\begin{figure}
  \centerline{\includegraphics[width=\textwidth]{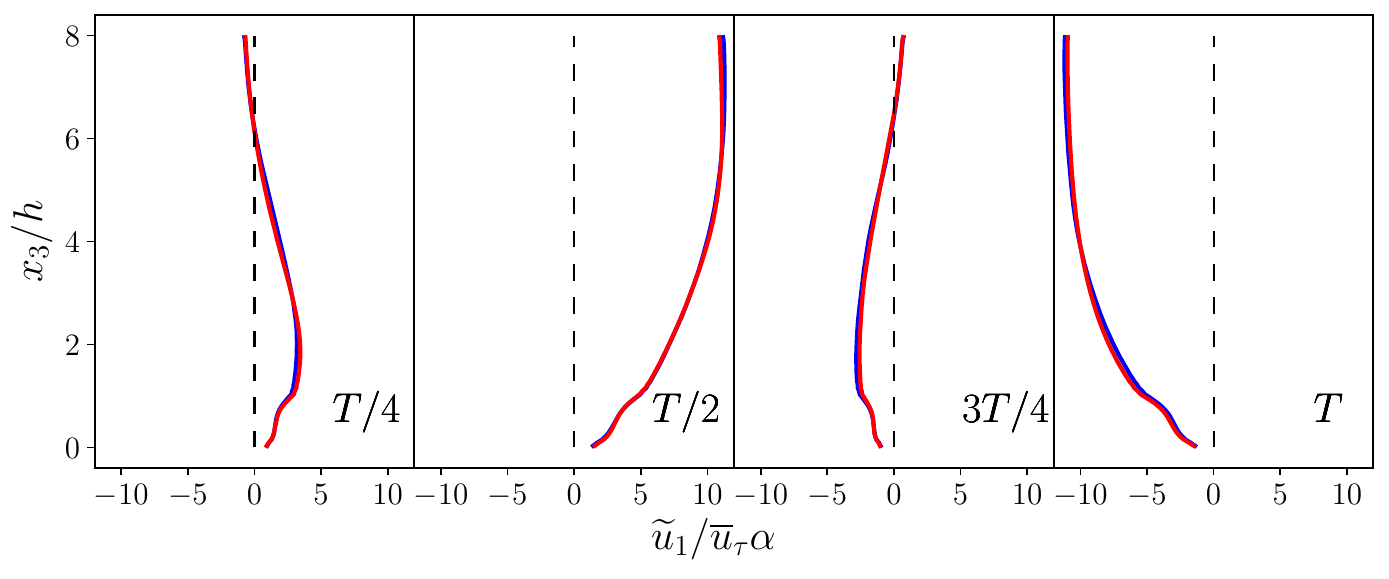}}
  \caption{Profiles of the oscillatory velocity $\widetilde{u}_1$ from the LL (blue) and HL (red) cases at different phases of the pulsatile cycle. Profiles are $T/4$ apart. }
\label{fig:osc_u_amp}
\end{figure}

\begin{figure}
  \centerline{\includegraphics[width=\textwidth]{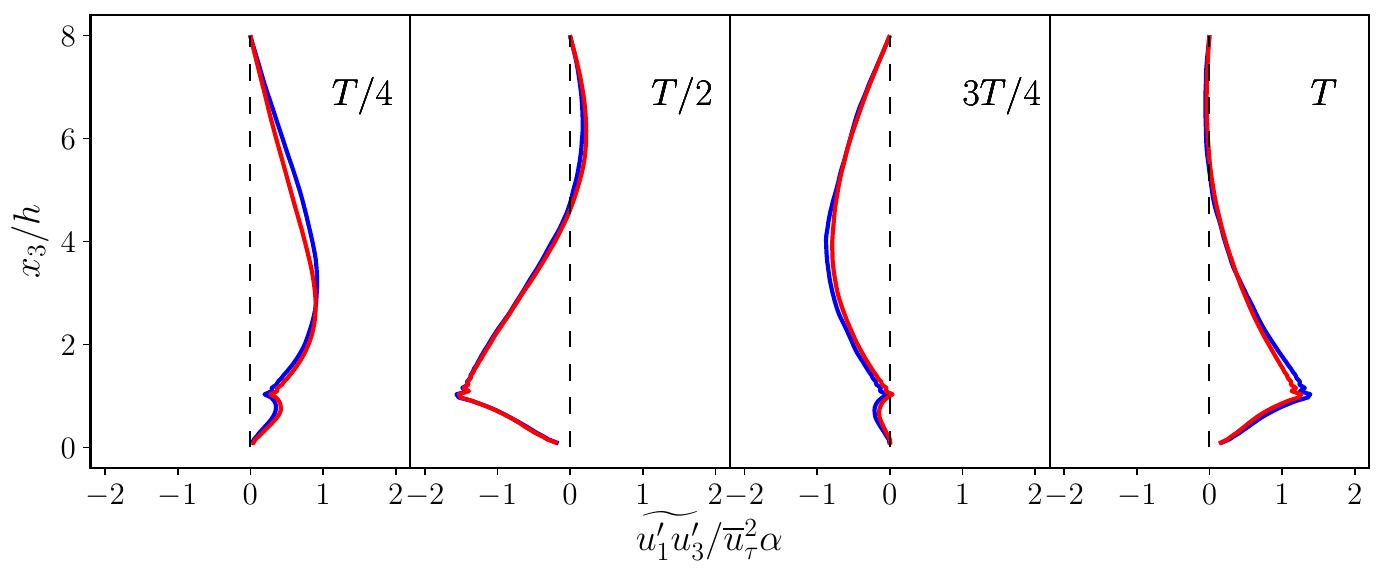}}
  \caption{Profiles of the oscillatory resolved Reynolds shear stress $\widetilde{u_1^\prime u_3^\prime}$ from the LL (blue) and HL (red) cases at different phases of the pulsatile cycle. Profiles are $T/4$ apart. }
\label{fig:osc_uw_amp}
\end{figure}

\begin{figure}
  \centerline{\includegraphics[width=\textwidth]{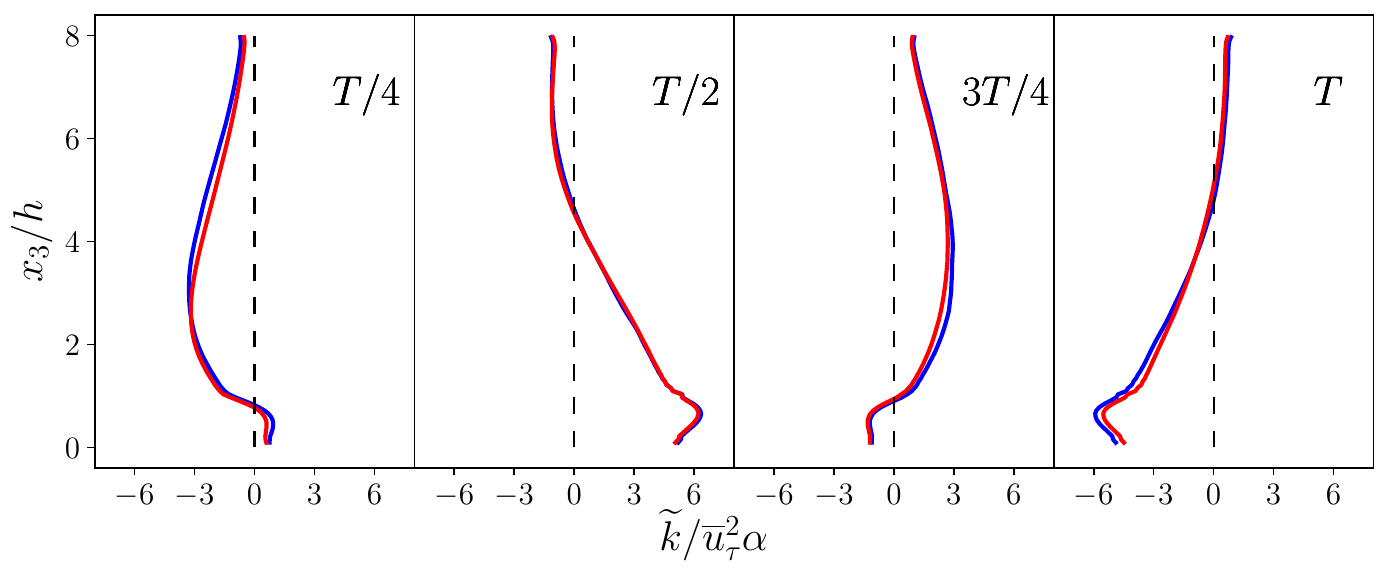}}
  \caption{Profiles of the oscillatory resolved turbulent kinetic energy $\widetilde{k}=(\widetilde{u_1^\prime u_1^\prime}+\widetilde{u_2^\prime u_2^\prime}+\widetilde{u_3^\prime u_3^\prime})/2$ from the LL (blue) and HL (red) cases at different phases of the pulsatile cycle. Profiles are $T/4$ apart.}
\label{fig:osc_tke_amp}
\end{figure}

\begin{figure}
  \centerline{\includegraphics[width=\textwidth]{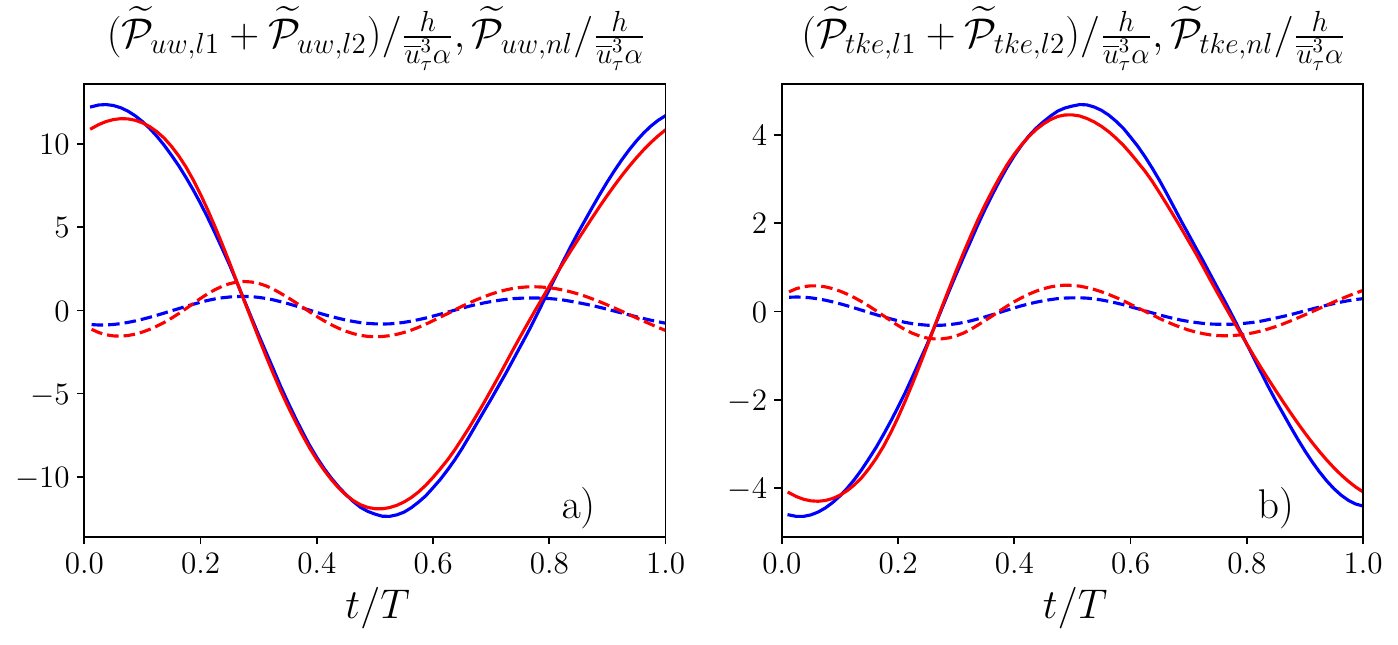}}
  \caption{Normalized production terms for $\widetilde{u_1^\prime u_3^\prime}$ (\textit{a}) and for $\widetilde{k}$ (\textit{b}) at $x_3/h=1.5$ from the LL (blue) and HL (red) cases. Solid lines denote the linear production terms, and dash lines represent the nonlinear production terms.}
\label{fig:tke_prod_decomp}
\end{figure}

Flow statistics from the LL and HL simulations are here examined to study the impact of the oscillation amplitude $(\alpha)$ on the oscillatory velocity and resolved Reynolds stresses. 
Only the LL and HL runs are discussed, as they were found to be representative of the observed behaviors for the other low and high-amplitude cases, respectively.

Figure \ref{fig:osc_u_amp} contrasts the profiles of oscillatory velocity $(\widetilde{u}_1)$ from the LL and HL runs at four equispaced phases during the pulsatile cycle.
$0<t<T/2$ and $T/2<t<T$ are flow acceleration and deceleration periods, respectively.
$\widetilde{u}_1$ at the top of the domain is controlled by the $\alpha$ parameter, so it is natural to use $\overline{u}_{\tau}\alpha$ as a normalization constant to study the problem. 
\cite{manna2012pulsating} investigated pulsatile open channel flow over an aerodynamically smooth surface, and showed that using such a normalization constant is indeed convenient as it leads to a collapse of $\widetilde{u}_1$ profiles across cases with different amplitudes, even in the presence of strong flow reversal.
This indicates that, at a given forcing frequency, the amplitude of the oscillatory velocity within the domain is proportional to that at the top of the domain.
In this work, we show that such a scaling works well also in the presence of aerodynamically rough surfaces, as evidenced by the excellent collapse of $\widetilde{u}_1/(\overline{u}_{\tau}\alpha)$  profiles in figure \ref{fig:osc_u_amp}.
This is not trivial, especially when considering the different scaling of surface drag between aerodynamically smooth and rough walls in the presence of flow unsteadiness.

The oscillatory resolved turbulent kinetic energy can be defined as
\begin{equation}
\widetilde{k}=\frac{1}{2}(\widetilde{u_1^\prime u_1^\prime}+\widetilde{u_2^\prime u_2^\prime}+\widetilde{u_3^\prime u_3^\prime})\ .
    \label{eq:tke_osc}
\end{equation}
As shown in figure \ref{fig:osc_uw_amp} and \ref{fig:osc_tke_amp}, both the oscillatory resolved Reynolds shear stress $\widetilde{u_1^\prime u_3^\prime}$ and $\widetilde{k}$ scale with $\overline{u}_{\tau}^2 \alpha$.
Although not shown, the three oscillatory normal Reynolds stresses also obey such a scaling, suggesting that any change in $\widetilde{k}$ is proportionally distributed to the three normal Reynolds stresses during the pulsatile cycle.
As discussed in the following paragraphs, the scaling of $\widetilde{u_1^\prime u_3^\prime}$ and $\widetilde{k}$ is a direct consequence of the mild non-linearity in the production of $\widetilde{k}$ and $\widetilde{u_1^\prime u_3^\prime}$.
Subtracting the budget equation of $\overline{u_1^\prime u_3^\prime}$ from that of $\langle u_1^\prime u_3^\prime \rangle$, one obtains,
\begin{equation}
    \frac{\partial \widetilde{u_1^\prime u_3^\prime}}{\partial t}= \underbrace{-2\widetilde{u_3^\prime u_3^\prime} \frac{\partial \overline {u }_1 }{\partial x_3}}_{\widetilde{\mathcal{P}}_{uw,l1}}
    \underbrace{-2\overline{u_3^\prime u_3^\prime} \frac{\partial \widetilde {u}_1}{\partial x_3}}_{\widetilde{\mathcal{P}}_{uw,l2}}
    \underbrace{-2\widetilde{u_3^\prime u_3^\prime} \frac{\partial \widetilde {u }_1 }{\partial x_3}+2\overline{\widetilde{u_3^\prime u_3^\prime} \frac{\partial \widetilde {u }_1 }{\partial x_3}}}_{\widetilde{\mathcal{P}}_{uw,nl}}+...\ ,
    \label{eq:bdg_uw_osc}
\end{equation}
where $\widetilde{\mathcal{P}}_{13,l1}$ and $\widetilde{\mathcal{P}}_{13,l2}$ are the linear production terms of $\widetilde{u_1^\prime u_3^\prime}$, while $\widetilde{\mathcal{P}}_{13,nl}$ is the nonlinear production. Similarly, the budget equations of $\widetilde{k}$ can be written as
\begin{equation}
    \frac{\partial \widetilde{k}}{\partial t}= \underbrace{-\widetilde{u_1^\prime u_3^\prime} \frac{\partial \overline {u }_1 }{\partial x_3}}_{\widetilde{\mathcal{P}}_{k,l1}}
    \underbrace{-\overline{u_1^\prime u_3^\prime} \frac{\partial \widetilde {u}_1}{\partial x_3}}_{\widetilde{\mathcal{P}}_{k,l2}}
    \underbrace{-\widetilde{u_1^\prime u_3^\prime} \frac{\partial \widetilde {u }_1 }{\partial x_3}+2\overline{\widetilde{u_1^\prime u_3^\prime} \frac{\partial \widetilde {u }_1 }{\partial x_3}}}_{\widetilde{\mathcal{P}}_{k,nl}}+...\ ,
    \label{eq:bdg_tke_osc}
\end{equation}
where $\widetilde{\mathcal{P}}_{k,l1}$ and $\widetilde{\mathcal{P}}_{k,l2}$ are the linear production terms of $\widetilde{k}$, while $\widetilde{\mathcal{P}}_{k,nl}$ is the nonlinear production.
As shown in figure \ref{fig:tke_prod_decomp}, the nonlinear production term is substantially smaller than the sum of the corresponding linear productions for both $\widetilde{u_1^\prime u_3^\prime}$ and $\widetilde{k}$.
Given that $\overline{u_1^\prime u_3^\prime}$, $\overline{u_3^\prime u_3^\prime}$, and ${\partial \overline{u}_1}/{\partial x_3}$ from the LL and HL cases are similar, when $\widetilde{u}_1 \sim \overline{u}_{\tau}\alpha$ and $\widetilde{u_3^\prime u_3^\prime}\sim \overline{u}_{\tau}^2 \alpha$, the total production of $\widetilde{u_1^\prime u_3^\prime}$ is
\begin{equation}
   \widetilde{\mathcal{P}}_{13}\approx \widetilde{\mathcal{P}}_{13,l1}+\widetilde{\mathcal{P}}_{13,l2}
   \sim \frac{\overline{u}_{\tau}^3 \alpha}{h}\ ,
    \label{eq:prod_uw_osc}
\end{equation}
whereas that of $\widetilde{k}$ is
\begin{equation}
   \widetilde{\mathcal{P}}_{k}\approx\widetilde{\mathcal{P}}_{k,l1}+\widetilde{\mathcal{P}}_{k,l2}
   \sim \frac{\overline{u}_{\tau}^3 \alpha}{h}\ .
    \label{eq:prod_k_osc}
\end{equation}
This in turn leads to the observed $\widetilde{u_1^\prime u_3^\prime} \sim \overline{u}_{\tau}^2 \alpha$ and $\widetilde{k} \sim \overline{u}_{\tau}^2 \alpha$.

This scaling is expected to fail under two conditions.
First, when $\alpha$ is sufficiently large, the contribution of $\widetilde{\mathcal{P}}_{k,nl}$ and $\widetilde{\mathcal{P}}_{k,nl}$ can no longer be neglected.
Second, when the variations in $\overline{u_1^\prime u_3^\prime}$ or $\overline{u_3^\prime u_3^\prime}$ among cases with different $\alpha$ are so large that the linear production terms do not scale with $\overline{u}_{\tau}^2 \alpha$ any more, which is what occurs within the UCL for the LVH and HVH cases (see figure \ref{fig:var_longtime}), (\ref{eq:prod_uw_osc}) and (\ref{eq:prod_k_osc}) do not hold.

The above analysis has shown that the $\alpha$ parameter primarily controls the amplitude of the oscillatory flow quantities, but has little impact on the wall-normal profiles of those quantities, which are instead controlled by $\omega$, as will be shown in the following section.
This behavior is also expected to hold in the smooth-wall setup \citep{manna2012pulsating}, although the mechanism responsible for generating drag over aerodynamically-smooth surfaces is quite distinct.

\subsubsection{Forcing frequency impacts on the oscillatory fields}\label{sec:osc_fields_freq}

This section discusses how the oscillatory velocity and resolved Reynolds stresses respond to variations in the forcing frequency.
Only low-amplitude cases will be considered since conclusions can be generalized across the considered runs.

\begin{figure}
  \centerline{\includegraphics[width=\textwidth]{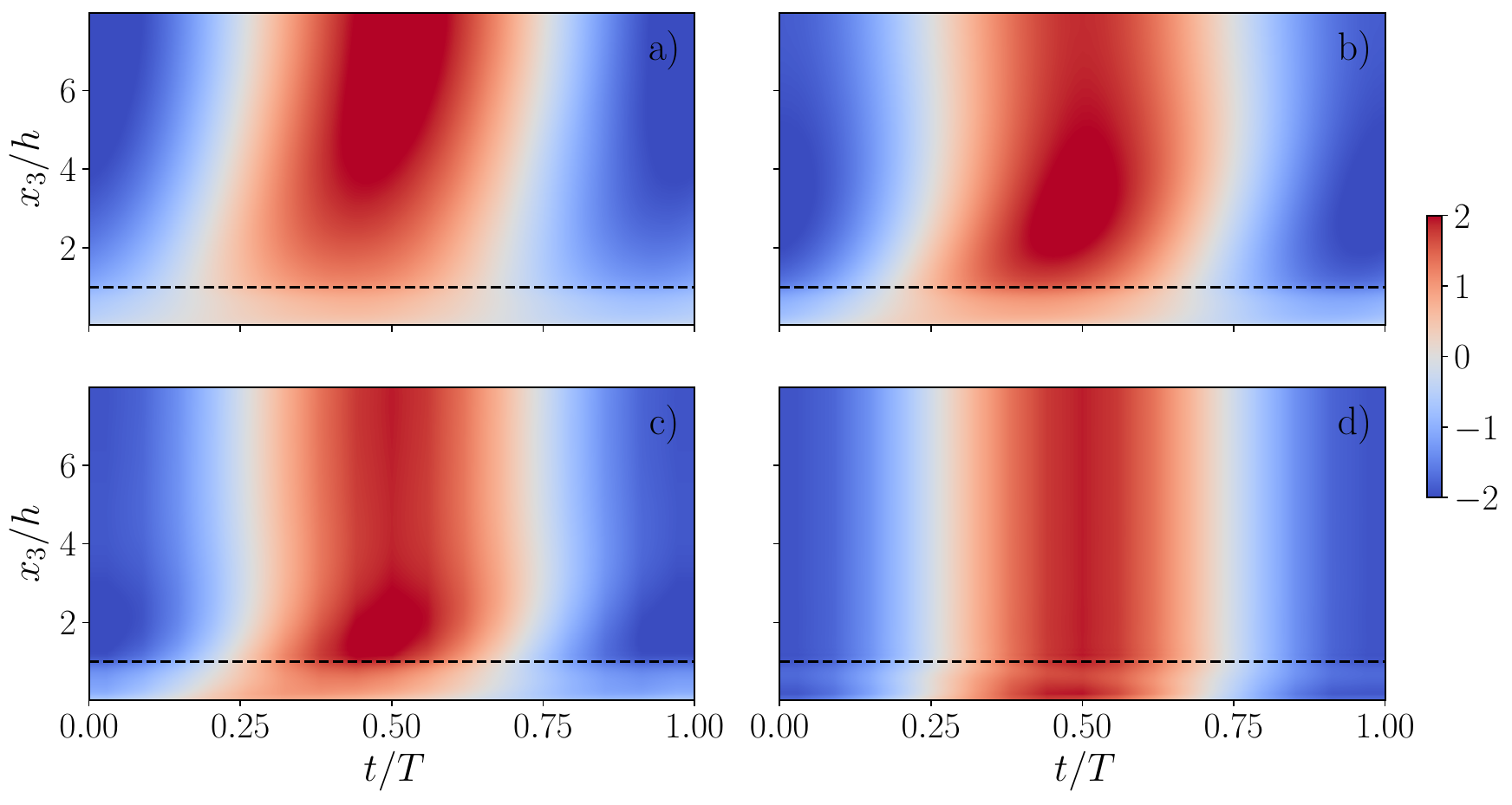}}
  \caption{Space–time diagrams of $\widetilde{u}_1/\overline{u}_\tau$ from the LL (\textit{a}), LM (\textit{b}), LH (\textit{c}), and LVH (\textit{d}) cases. Horizontal dashed lines identify the top of the UCL.}
\label{fig:contour_u}
\end{figure}

\begin{figure}
  \centerline{\includegraphics[width=\textwidth]{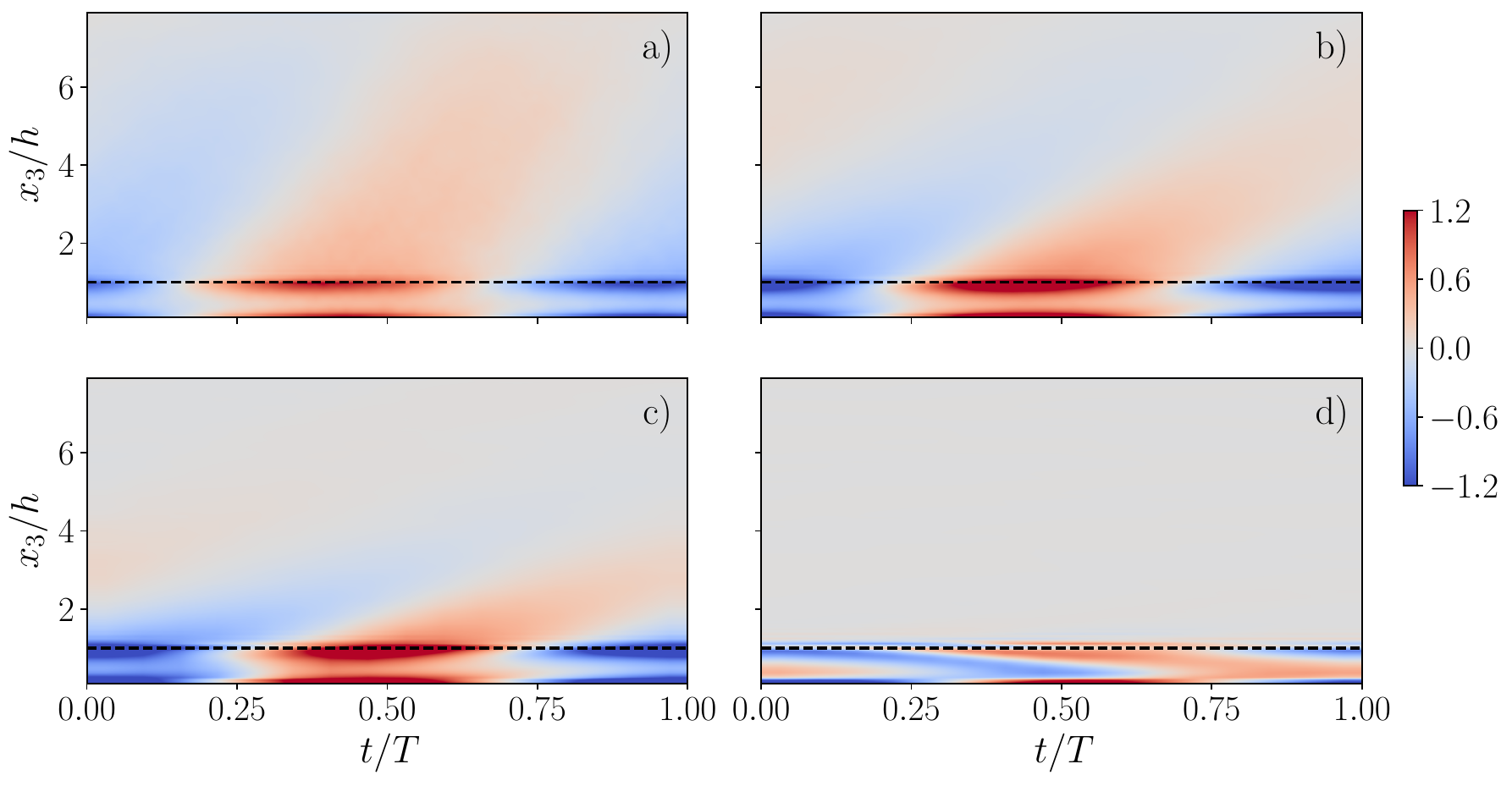}}
  \caption{Space–time diagrams of $(h/\overline{u}_\tau) {\partial \widetilde{u}_1}/{\partial x_3}$ from the LL (\textit{a}), LM (\textit{b}), LH (\textit{c}), and LVH (\textit{d}) cases.}
\label{fig:contour_dudz}
\end{figure}

Figure \ref{fig:contour_u} and \ref{fig:contour_dudz} present the time evolution of $\widetilde{u}$ and $\partial \widetilde{u}/\partial x_3$.
Three distinct frequency regimes can be identified.
The first regime corresponds to the highest amongst the considered forcing frequencies, i.e., the LVH case.
For this flow regime, the oscillation in $\partial \widetilde{u}/\partial x_3$ is typically confined within the UCL.
This behavior can be best explained when considering that the time period of the oscillation is comparable to the eddy turnover time of turbulence in the UCL, i.e., $T \approx T_h$, which is the characteristic time scale for ``information transport" within the UCL.
At the three lower forcing frequencies, i.e., the LL, LM, and LH cases, on the contrary, the interaction between the roughness elements and the unsteady flow induces an oscillation in the shear rate, which has a phase lag of roughly $\pi/2$ with respect to the pulsatile forcing at the top of the UCL.
This oscillating shear rate then propagates in the positive wall-normal direction while being progressively attenuated.
The propagation speed of the oscillating shear rate appears to be constant for a given forcing frequency, which can be readily inferred by the constant tilting angle in the $\partial \widetilde{u}_1/\partial x_3$ contours.
The flow region affected by the oscillating shear rate defines the so-called ``Stokes layer".
For cases with two moderate frequencies, i.e., the LM and LH cases, the Stokes layer thickness $(\delta_s)$ is smaller than the domain height $L_3$.
Above the Stokes layer, the slope of $\widetilde{u}_1$ is nominally zero over the pulsatile cycle, and the flow in such a region resembles a plug-flow.
On the contrary, in the LL case, the entire domain is affected by the oscillating shear rate, thus $\delta_s>L_3$.

\begin{figure}
  \centerline{\includegraphics[width=\textwidth]{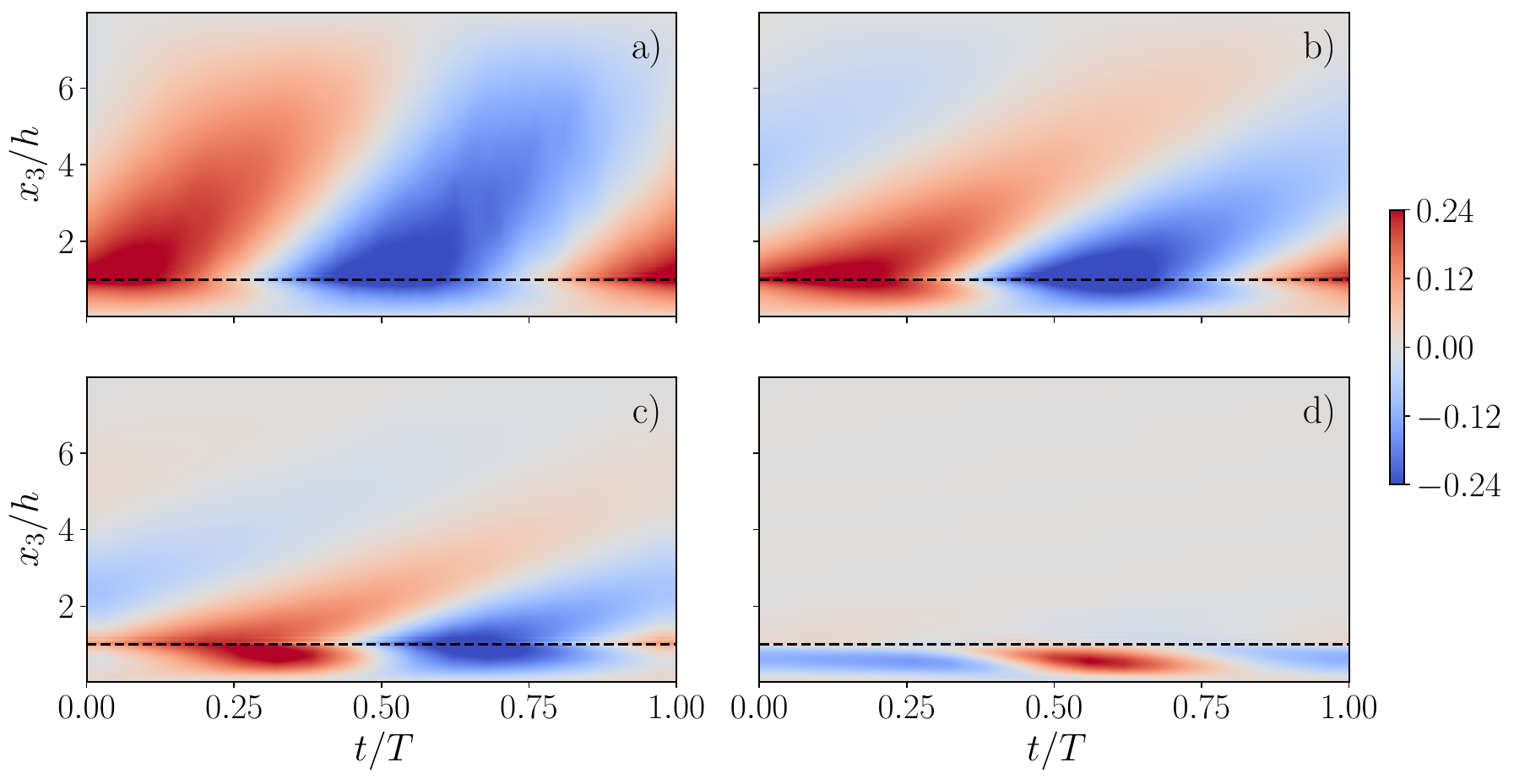}}
  \caption{Space–time diagrams of $\widetilde{u_1^\prime u_3^\prime} / \overline{u}_\tau^2$ from the LL (\textit{a}), LM (\textit{b}), LH (\textit{c}), and LVH (\textit{d}) cases.}
\label{fig:contour_uw}
\end{figure}
\begin{figure}
  \centerline{\includegraphics[width=\textwidth]{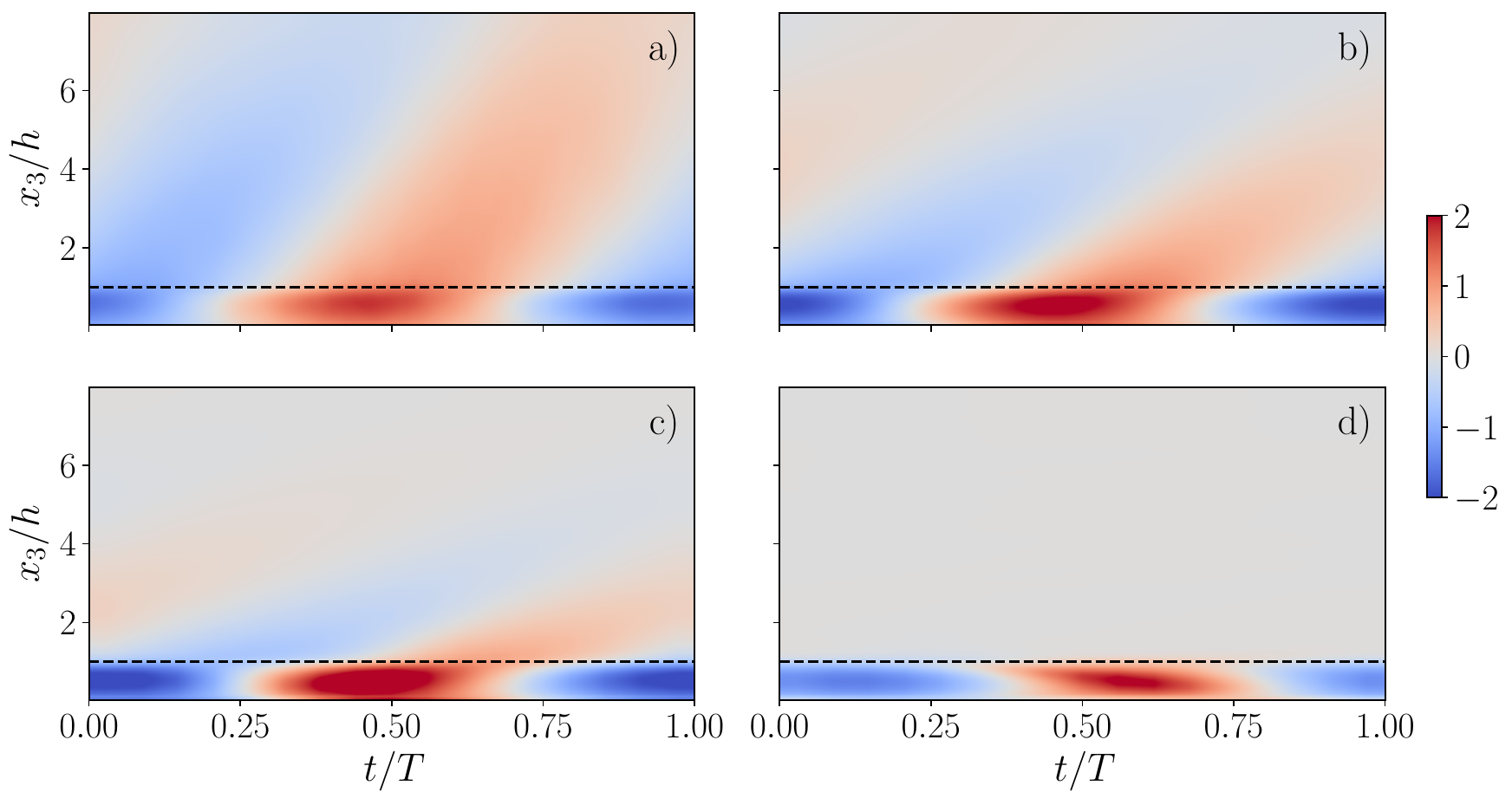}}
  \caption{Space–time diagrams of $\widetilde{u_1^\prime u_1^\prime}/\overline{u}_\tau^2$ from the LL (\textit{a}), LM (\textit{b}), LH (\textit{c}), and LVH (\textit{d}) cases.}
\label{fig:contour_uu}
\end{figure}

Figure \ref{fig:contour_uw} and \ref{fig:contour_uu} depict the time evolution of $\widetilde{u_1^\prime u_3^\prime}$ and $\widetilde{u_1^\prime u_1^\prime}$, respectively.
Although the contours of $\widetilde{u_2^\prime u_2^\prime}$ and $\widetilde{u_3^\prime u_3^\prime}$ are not shown, these quantites vary in a similar fashion as $\widetilde{u_1^\prime u_1^\prime}$ during the puslatile cycle.
These space–time diagrams confirm that the considered frequencies encompass three distinct flow regimes.
For the LVH case, time variations of the oscillatory resolved Reynolds stresses are essentially zero above the UCL.
In cases with three lower frequencies, oscillatory resolved Reynolds stresses exhibit a similar behavior to $\partial \widetilde{u}_1/\partial x_3$.
Specifically, there appear oscillating waves propagating away from the UCL at a constant speed and meanwhile getting weakened.
In the LM and LH cases, such oscillating waves are fully dissipated at the upper limit of the Stokes layer, above which the turbulence is ``frozen" and passively advected.

\begin{figure}
  \centerline{\includegraphics[width=\textwidth]{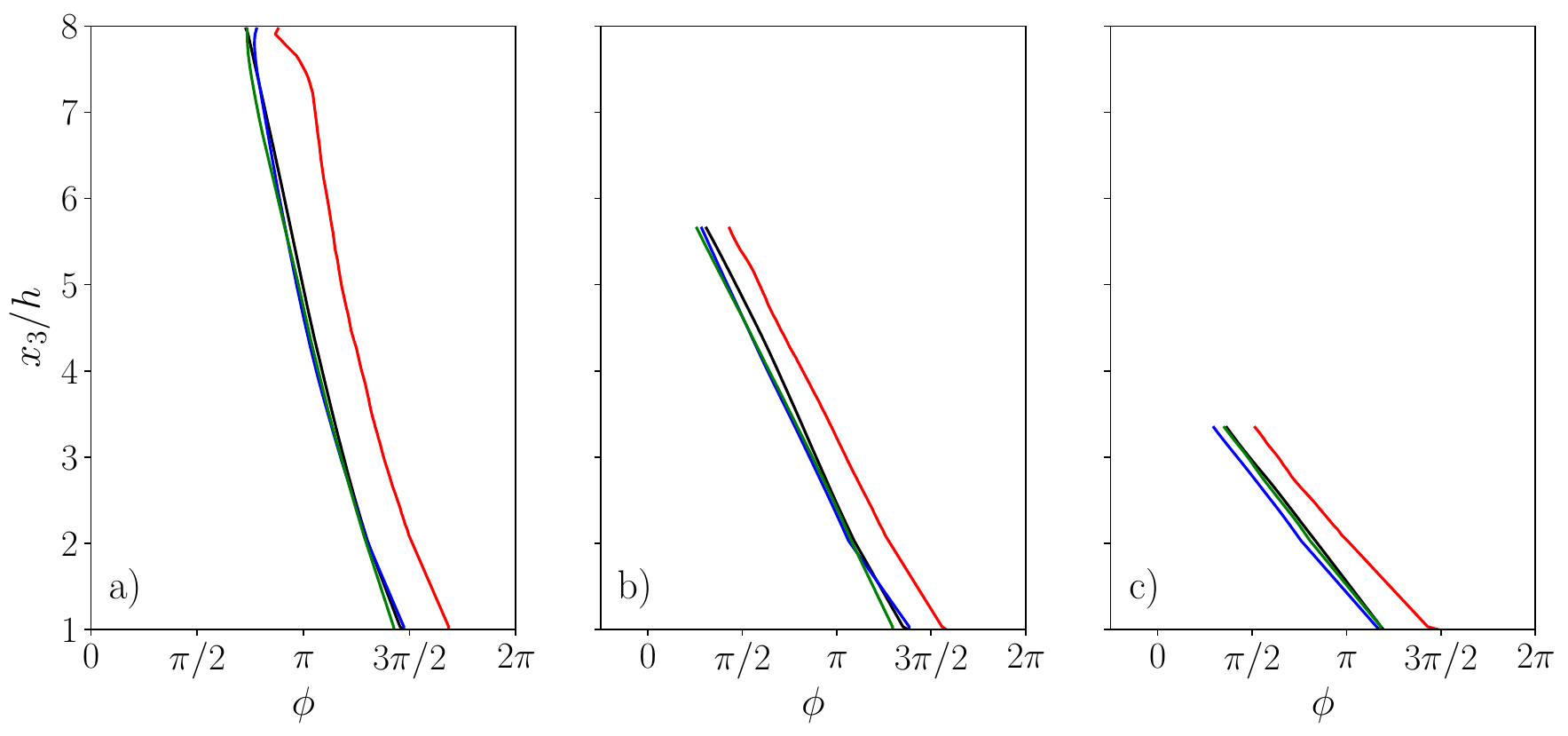}}
  \caption{Phase lag of ${\partial \widetilde{u}_1}/{ \partial x_3}$ (red), $-\widetilde{ u_1^\prime u_3^\prime }$ (black), $\widetilde{ u_1^\prime u_1^\prime }$ (magenta), $\widetilde{ u_2^\prime u_2^\prime }$ (green), and $\widetilde{  u_3^\prime u_3^\prime}$ (blue) with respect to the pulsatile forcing from the LL (left), LM (middle), and LH (right) cases.}
\label{fig:phase_lag_var}
\end{figure}

A visual comparison of the tilting angles in the contours of oscillatory resolved Reynolds stresses and $\partial \widetilde{u}_1/\partial x_3$ reveals that the oscillating waves in these quantities feature similar propagation speeds.
This behavior closely resembles the one observed in smooth-wall cases \citep{scotti2001numerical,manna2015pulsating}.
The physical interpretation is that when the oscillating shear rate is diffused upwards by the background turbulent flow, it interacts with the local turbulence via the mechanisms described in (\ref{eq:bdg_uw_osc}) and (\ref{eq:bdg_tke_osc}), thus inducing the observed oscillations in the resolved Reynolds stresses.
To further quantify the propagation speeds of the oscillating waves in ${\partial \widetilde{u}_1}/{ \partial x_3}$ and oscillatory resolved Reynolds stresses, figure \ref{fig:phase_lag_var} present the phase lag of those quantities with respect to the pulsatile forcing.
For a quantity $\theta$, the propagation speed $c_\theta$ is defined based on the slope of the phase lag, i.e.,
\begin{equation}
    c_\theta=- \omega\frac{\partial x_3}{\partial \phi_\theta}\ .
    \label{eq:wave_speed}
\end{equation}

Table~\ref{tab:wave_speed} summarizes the wave propagation speeds for $\partial \widetilde{u}_1 /\partial x_3$, $-\widetilde{ u_1^\prime u_3^\prime }$, $\widetilde{ u_1^\prime u_1^\prime }$, $\widetilde{ u_2^\prime u_2^\prime }$, and $\widetilde{  u_3^\prime u_3^\prime}$ of each case.
This again confirms that the oscillating waves propagate at a similar speed for the considered quantities.
It is also noteworthy to point out that the speed of the propagating wave increases with $\omega$.

\begin{table}
  \begin{center}
\def~{\hphantom{0}}
  \begin{tabular}{lccccc}
        Case  & $c_{du1dx3}/\overline{u}_\tau$ & $c_{-u1u3}/\overline{u}_\tau$ & $c_{u1u1}/\overline{u}_\tau$ & $c_{u2u2}/\overline{u}_\tau$ & $c_{u3u3}/\overline{u}_\tau$\\[3pt]
      LL   & 0.50 & 0.51 & 0.51 & 0.49 & 0.48\\
      LM   & 0.55 & 0.54 & 0.60 & 0.57 & 0.58 \\
      LH   & 0.72 & 0.71 & 0.77 & 0.76 & 0.74 \\
  \end{tabular}
  \caption{Propagation speeds of oscillating waves in $\partial \widetilde{u}_1 /\partial x_3$ and oscillatory resolved Reynolds stresses.}
  \label{tab:wave_speed}
  \end{center}
\end{table}

Three other observations can be made from figure \ref{fig:phase_lag_var}.
First, throughout the considered cases, there appears a marked phase lag of roughly $\pi/6$ between $\partial \widetilde{u}_1 /\partial x_3$ and (negative) $\widetilde{u_1^\prime u_3^\prime}$, indicating a deviation from the Boussinesq eddy viscosity assumption.
\cite{weng2016numerical} reported a similar finding, and they attributed such behavior to non-equilibrium effects arising when the time period of the pulsatile forcing is short compared to the local turbulence relaxation time so that the turbulence is not able to relax to its equilibrium state during the pulsation cycle.
Second, the lack of phase lag among the oscillatory normal resolved Reynolds stresses implies that the oscillatory pressure-redistribution terms respond immediately to the change in $\widetilde{u_1^\prime u_1^\prime}$.
Third, the lifetimes of oscillating waves in the resolved Reynolds stresses and shear rate, which is inferred by the difference between the phase lags at the top of the UCL and at the upper limit of the Stokes layer, are no more than half of the oscillation time period, although they decrease with $\omega$.
They are considerably shorter than those in smooth-wall cases, which are typically larger than one oscillation period \citep{scotti2001numerical,manna2015pulsating,weng2016numerical}.

\subsubsection{Scaling of the Stokes layer thickness}\label{sec:scaling_delta_s}

$\delta_s$ is a quantity of interest across many applications, since it defines the region where the turbulence and the mean flow are out of equilibrium.
In such a region, established turbulence theories may fail to capture flow dynamics that are of relevance for, e.g., surface drag and scalar dispersion.

For the wave-current boundary flow, where the surface is typically transitionally rough, the wave boundary layer thickness---an equivalent of Stokes layer thickness---scales as
\begin{equation}
    \delta_w \sim \frac{\kappa u_{\tau,max}}{\omega}\ ,
    \label{eq:delta_wave_bl}
\end{equation}
where $u_{\tau,max}$ is the friction velocity based on the maximum phase-averaged wall stress during the pulsatile cycle \citep{grant1979combined}. 
Such a scaling argument is not valid for the current cases, even though the considered surface is also rough.
As shown in \S\ref{sec:osc_fields_amp}, normalized oscillatory velocity and resolved Reynolds stresses profiles collapse between cases with the same frequencies, implying that the Stokes layer thickness is only dependent on $\omega$, whereas $\tau_{max}$ is determined by both $\alpha$ and $\omega$.
Rather, the scaling of $\delta_s$ in the current cases is a trivial extension of the model first introduced by \cite{scotti2001numerical}, as discussed next.

Let us recall from \S\ref{sec:intro} that the Stokes layer thickness for turbulent pulsatile flow over aerodynamically-smooth surface  \citep{scotti2001numerical} is defined as
\begin{equation}
    \delta_s = 2\frac{\kappa \overline{u}_\tau}{\omega}(1+\sqrt{1+ \frac{2\nu \omega}{\kappa^2 \overline{u}_\tau^2}})\ .
    \label{eq:delta_s_pio_2}
\end{equation}

Here we apply two modifications to this model in order to make it applicable to the current rough-wall cases.
First, given that the viscous stress is omitted, the molecular viscosity $\nu$ can be neglected.
Also, in the current cases, the oscillating shear rate is generated within the UCL rather than at the bottom surface (as in the smooth-wall cases), and the extent of the oscillating shear rate propagation defines the thickness of the Stokes layer.
This behavior can be easily captured by augmenting $\delta_s$ by the displacement height ($d$).
Specifically, we draw an analogy to smooth-wall cases by taking $d$ as the offset, since it is the virtual origin of the longtime-averaged velocity profile. 
$d$ is a plausible choice of the offset since it captures the limiting behavior of the flow system as the canopy packing density varies. 
For instance, in the limit of $\lambda_p \rightarrow 0$ (very sparse canopies), $d = 0$, i.e., the oscillating shear rate grows from the bottom of the domain. 
On the contrary, in the limit of $\lambda_p \rightarrow 1$ (very dense canopies), $d = h$, i.e., the oscillating shear layer starts at the top of the UCL.
Based on these considerations, a phenomenological model for the Stokes layer thickness is 
\begin{equation}
    \delta_s=\frac{4\kappa u_\tau}{\omega}+d\ .
    \label{eq:delta_s}
\end{equation}
Note that, in the limit of $\omega \rightarrow 0$, the Stokes layer no longer exists, rendering (\ref{eq:delta_s}) invalid.
At this limit, as previously stated in \cite{scotti2001numerical}, $T$ is much larger than the turbulence relaxation time.
 As a result, the turbulence maintains a quasi-equilibrium state, and the flow statistics are indistinguishable from those of the corresponding equilibrium canopy layer flows, if scaled with the instantaneous inner/outer units.

\begin{figure}
  \centerline{\includegraphics[width=\textwidth]{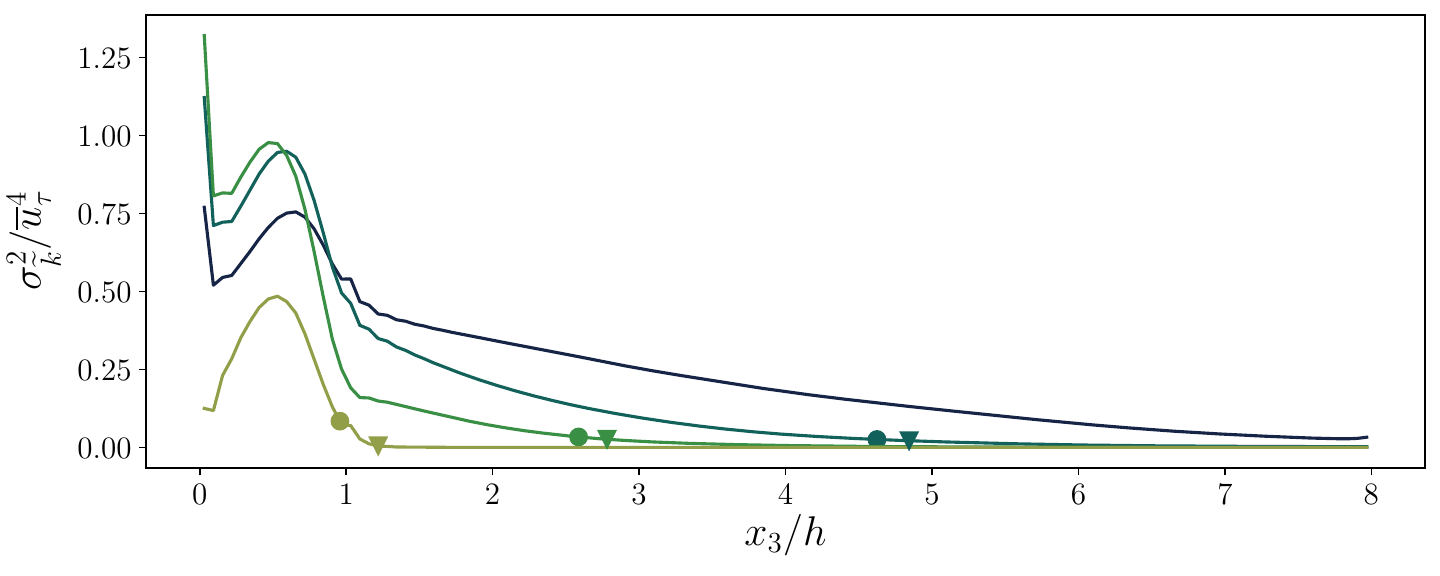}}
  \caption{Time variance of $\widetilde{k}$: $\sigma^2_{ \widetilde{k}}=\int^{T}_0 (\frac{1}{2}(\widetilde{ u_1^\prime u_1^\prime }+\widetilde{ u_2^\prime u_2^\prime }+\widetilde{ u_3^\prime u_3^\prime }))^2 dt/T$ of the LL (navy blue), LM (dark green), LH (light green), and LVH (yellow green) cases. Circle denotes $\delta_s$ estimated via (\ref{eq:delta_s}). Triangle represents the location where $\sigma^2_{ \widetilde{k}}$ is reduced to $1\%$ of its maximum value.}
\label{fig:varian_tke}
\end{figure}

Figure \ref{fig:varian_tke} compares the predictions of (\ref{eq:delta_s}) against LES results.
Note that only low-amplitude cases are shown, since, as mentioned earlier, $\delta_s$ only depends on $\omega$.
The upper limit of the Stokes layer is identified as the location where $\sigma^2_{ \widetilde{k}}$ is $1\%$ of its maximum, where
\begin{equation}
   \sigma^2_{ \widetilde{k}}= \frac{1}{T}\int^{T}_0 (\frac{1}{2}(\widetilde{ u_1^\prime u_1^\prime }+\widetilde{ u_2^\prime u_2^\prime }+\widetilde{ u_3^\prime u_3^\prime }))^2 dt\ .
    \label{eq:sigma2}
\end{equation}
is the time variance of $\widetilde{k}$.
From figure \ref{fig:varian_tke}, it is apparent that the estimated $\delta_s$ compare very well with LES results.
The estimation of $\delta_s$ for the LL case is not shown in figure \ref{fig:varian_tke} because it exceeds the height of the computational domain.
(\ref{eq:delta_s}) can hence be used in future studies to identify the Stokes layer thickness for pulsatile flows over aerodynamically rough surfaces.

\section{Conclusions}\label{sec:conclusion}
This paper has examinined the impact of flow pulsation on longtime- and phase-averaged flow statistics in an open-channel flow over urban-like roughness.
A series of LES of pulsatile flow past an array of cuboid elements has been carried out, programmatically varying the oscillation amplitude $(\alpha)$ and frequency $(\omega)$.
The forcing frequencies have been chosen as a multiple of the characteristic frequency of turbulence in the UCL and encompass a range of values representative of submesoscale motions \citep{mahrt2020non}. 
The main findings and contributions of this study are outlined below.

\begin{enumerate}
\item Flow pulsation leads to an increase of the $z_0$ parameter educed from longtime-averaged $\overline{u}_1$ profiles, with larger $\alpha$ and $\omega$ values yielding a larger $z_0$.
On the contrary, $d$ was found to be insensitive to variations in $\alpha$ and $\omega$.
The increase of $z_0$ was shown to be a direct consequence of the quadratic relation between the phase-averaged canopy drag $\langle D \rangle$ and the phase-averaged velocity $\langle u_1\rangle$, and this relation was leveraged to construct a phenomenological model for $z_0$. 
The proposed model takes surface information and the variance of the phase-averaged velocity in the UCL as input parameters and captures the impact of flow unsteadiness on the $z_0$ parameter in the absence of flow reversal.
\item The wall-normal distributions of the longtime-averaged shear stress and canopy drag are unaltered by the flow unsteadiness. In contrast, the same cannot be said for longtime-averaged resolved normal Reynolds stresses, especially in the UCL.
In particular, $\overline{k}$ profiles were found to be relatively more sensitive to variations in $\alpha$ via the longtime-averaged shear production of $\overline{k}$.
The highest frequency cases were also characterized by a relatively more isotropic turbulence field in the UCL, owing to a more efficient kinetic energy redistribution by the pressure-strain terms.
\item The oscillation amplitudes of phase-averaged streamwise velocity and resolved Reynolds stresses scale with $\alpha$.
This behavior is due to the fact that the nonlinear production terms of $\widetilde{u_1^\prime u_3^\prime}$ and $\widetilde{k}$ are of relatively modest magnitude when compared to the linear ones. Increasing the pulsation amplitude might lead to more substantial contributions from nonlinear production terms and break down this scaling. 
\item For each case, profiles of oscillatory shear rate and resolved Reynolds stresses are characterized by oscillating waves which are advected away from the UCL at a constant speed while also being dissipated.
$\omega$ is found to determine both the speeds of the oscillating waves and the extent of these waves, which identifies the Stokes layer thickness $(\delta_s)$.
More specifically, $\delta_s$ was found to increase with decreasing $\omega$, whereas the wave speed increased with $\omega$.
The scaling of $\delta_s$ has also been discussed, and findings have been used to propose a model for $\delta_s$.
\end{enumerate}

All in all, flow pulsation is found to have a significant impact on both longtime-averaged and phase-averaged flow statistics, with nuanced dependencies on oscillation amplitude and frequency.
The observed enhancement of the longtime-averaged surface drag, the isotropization of turbulence in the UCL, and the presence of a Stokes layer, amongst others, are expected to have important implications on the exchange of mass, energy, and momentum between the land surface and the atmosphere, as well as affect our ability to model these processes in weather forecasting and climate models. 
These models typically rely on surface flux parameterizations and theories that are based on flow stationarity assumptions and are not able to capture these behaviors correctly \citep[see, e.g.,][]{Stensrud2007}.
The proposed phenomenological models for $z_0$ and $\delta_s$, as well as the identified scaling of phase-averaged flow statistics, contribute to advancing our understanding of flow unsteadiness in the ABL and offer a pathway for the development of improved surface flux parameterizations.  
Given the massive parameter space of unsteady ABL flow processes, it is also essential to acknowledge that several questions remain unanswered and deserve further investigation.
For example, what is the impact of different types of periodic and aperiodic flow unsteadiness on turbulence statistics and topology?
How are these variations in the structure of turbulence impacting land-atmosphere exchange rates of momentum, energy, and mass?
Can prevailing surface flux parameterizations be modified to account for these impacts?
Addressing these questions will be the subject of future studies.\\

\noindent \textbf{Declaration of Interests.} The authors report no conflict of interest. \\

\noindent \textbf{Acknowledgements.} The authors acknowledge support from the Department of Civil Engineering and Engineering Mechanics at Columbia University. 
This material is based upon work supported by, or in part by, the Army Research Laboratory and the Army Research Office under contract/grant number W911NF-22-1-0178. 
This work used the Stampede2 cluster at the Texas Advanced Computing Center through allocation ATM180022 from the Extreme Science and Engineering Discovery Environment (XSEDE), which was supported by National Science Foundation grant number \#1548562.

\appendix
\section{Wall-layer modeling considerations}\label{sec:wall_model_justify}

As discussed in \S\ref{sec:meth}, simulations have been conducted using an algebraic wall-layer model at the solid-fluid interface to evaluate tangential surface stresses. 
In this section, we show that the use of an equilibrium wall-layer model can be justified on the basis that the flow is in the fully rough aerodynamic regime.

An LES of flow over a single cube is carreid out at roughness Reynolds number $Re_{\tau}=\overline{u}_{\tau}h/\nu=400$ (hereafter referred as to LES400), and results are compared with those from a DNS run (DNS400).
At such a Reynolds number, the flow field is in fully rough regime, as also shown in \cite{xie2008large}.
The size of the computational domain is $[0,3h] \times [0,3h] \times [0,4h]$, and the planar and frontal area densities are the same as those in the main simulations of the study. 
The forcing frequency and the oscillation amplitude are $\omega T_h=0.125\pi$ and $\alpha=0.8$, respectively, which are comparable to the ones considered in the study. 
The grid resolution of LES400 follows the main simulations, which is $(n_1, n_2, n_3) = (8, 8, 16)$ for each cube, and the identical wall-layer model as that in the main simulations is applied in the vicinity of the cube facets and the lower surface with the same roughness length scale.
The grid resolution of DNS400 is $(n_1, n_2, n_3) = (64, 64, 128)$ per cube.
Such a grid resolution ensures that the ratio between the grid size $\Delta=\sqrt[3]{\Delta_1 \Delta_2 \Delta_3}$ and the Kolmogorov scale $\eta$ does not exceed 2, which has been proven sufficient for DNS of flow over fully rough surfaces \citep{zhang2022evidence}.

\begin{figure}
  \centerline{\includegraphics[width=\textwidth]{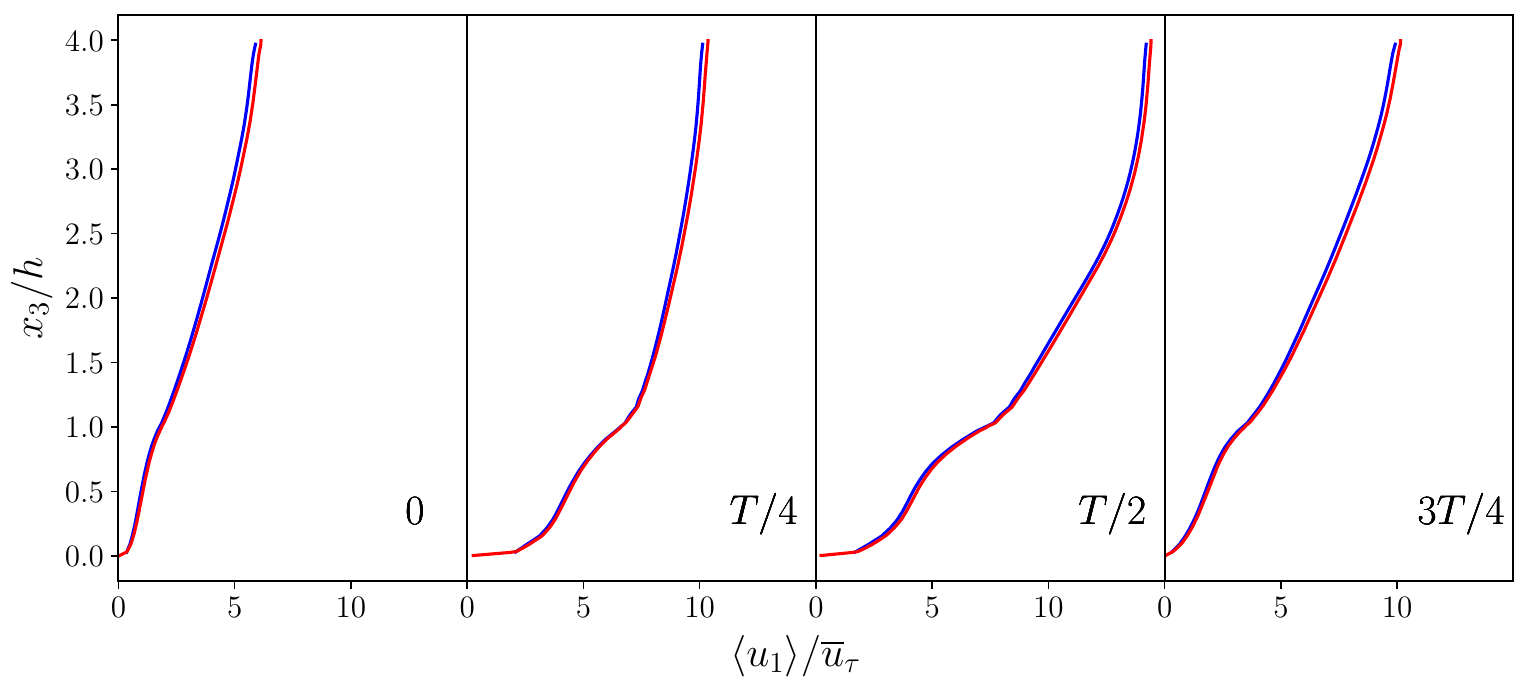}}
  \caption{Phase-averaged velocity $\langle u_1 \rangle$ of LES400 (blue) and DNS400 (red).}
\label{fig:small_domain_u}
\end{figure}

\begin{figure}
  \centerline{\includegraphics[width=\textwidth]{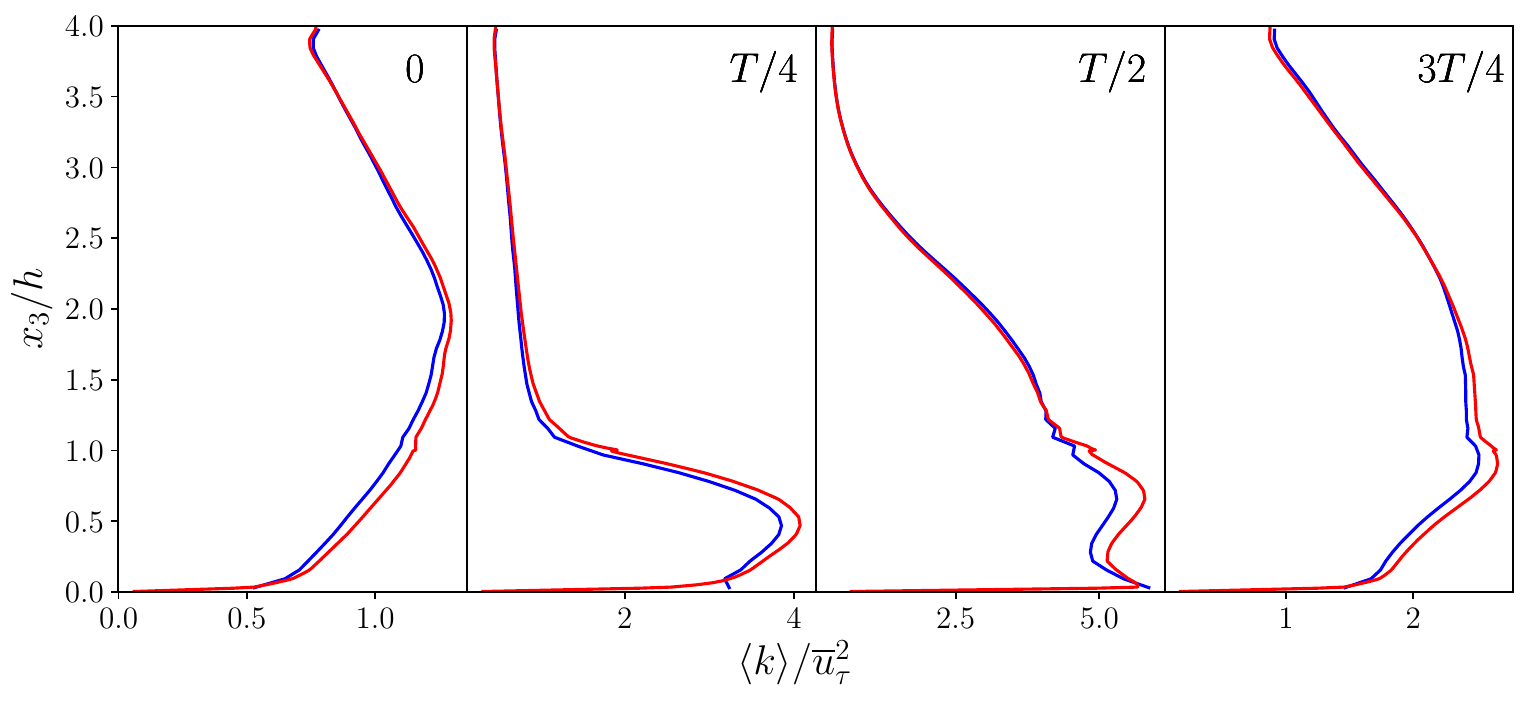}}
  \caption{Phase-averaged turbulent kinetic energy $\langle k \rangle = \langle u_i^\prime u_i^\prime \rangle /2$ of LES400 (blue) and DNS400 (red). }
\label{fig:small_domain_k}
\end{figure}

In both simulations, the contribution from the tangential stresses at the cube facets and lower surface to the total surface drag remains below $1\%$, confirming that the flow is in the fully rough regime.
Figure~\ref{fig:small_domain_u} and \ref{fig:small_domain_k} display the phase-averaged velocity $(\langle u_1 \rangle)$ and turbulent kinetic energy ($\langle k \rangle / \overline{u}_\tau^2$), respectively. 
Profiles from the LES400 case are in good agreement with corresponding DNS quantities, with the maximum error in the LES400 profiles relative to those from the DNS400 case being approximately 1\% and 6\% for $(\langle u_1 \rangle)$ and $\langle k \rangle$, respectively. 
The minor mismatches in $\langle k \rangle$ can be partly explained by the fact that the SGS contribution to $\langle k \rangle$ is zero for the LES400 case.
It is suggested that, although the equilibrium assumption does not hold in a strict sense, the use of an equilibrium wall-layer model does not result in a noticeable impact on model results for the considered unsteady flow cases.

\section{Resolution sensitivity analysis}\label{sec:gird_sensitivity}

To identify grid resolution requirements for simulations in this study, a grid-resolution sensitivity analysis has been conducted for the stationary flow case, i.e., $\alpha=0$.
The domain size for this analysis is $(36h, 12h, 4h)$, and we have studied the convergence of $\overline{u}_1$, $\overline{u_1^\prime u_1^\prime}$, and $\overline{u_1^\prime u_3^\prime}$ profiles as the grid stencil is progressively reduced. 
Three grid resolutions have been considered, namely $(4, 4, 8)$, $(8, 8, 16)$, and $(12, 12, 24)$ on a per-cube basis.
Note that the reduced domain size may have an impact on the evaluated flow statistics, but this serves the purpose of this analysis, since we are here only interested in quantifying relative variations of selected profiles across grid resolutions.
Other numerical and physical parameters of the grid-sensitivity analysis simulations are set equal to the ones used in the main simulations.

\begin{figure}
  \centerline{\includegraphics[scale=0.45]{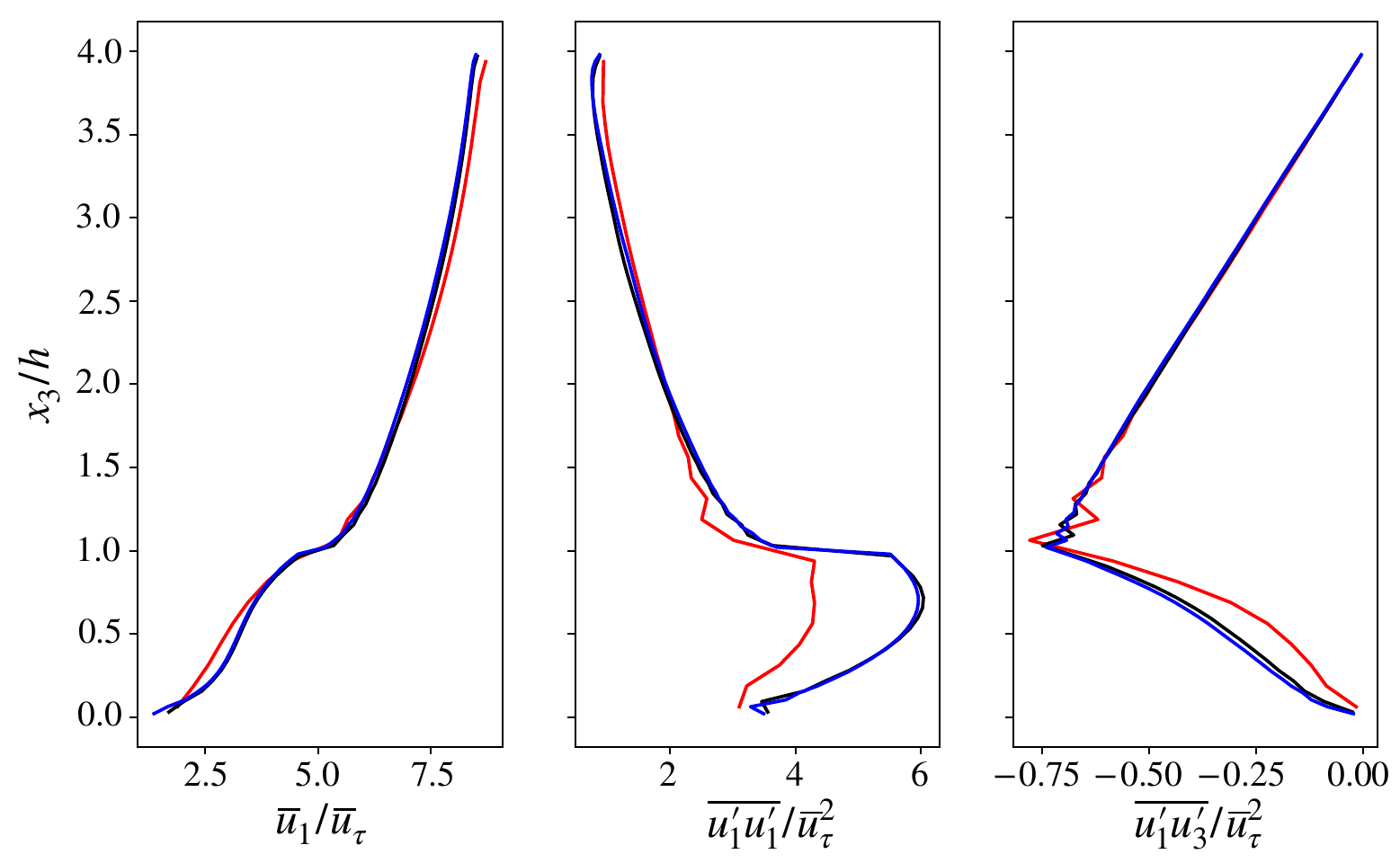}}
  \caption{Comparison of the mean flow, resolved streamwise velocity variance, and resolved Reynolds shear stress of the three test simulations for the resolution sensitivity analysis. Red: $(4, 4, 8)$; black: $(8, 8, 16)$; blue: $(12, 12, 24)$.}
\label{fig:grid_sensitivity}
\end{figure}

As apparent from figure \ref{fig:grid_sensitivity}, profiles from the $(8, 8, 16)$ case are essentially matching corresponding ones from the $(12, 12, 24)$ case, indicating that the chosen grid resolution for the pulsatile channel flow analysis is sufficient to yield grid-independent flow statistics (up to second order).

\bibliographystyle{jfm}

\bibliography{main}

\end{document}